\documentclass[final,1p,times]{elsarticle}
\usepackage{amsmath}
\usepackage{url}
\usepackage{hyphenat}
\usepackage{graphicx}
\usepackage{dcolumn}
\usepackage{bm}

\def\eqref#1{(\ref{#1})}
\def\l{\langle}

\def\r{\rangle}

\newcolumntype{t}[1]{D{.}{.}{#1}}
\newcolumntype{.}{D{.}{.}{-1}}

\biboptions{square,comma,numbers,sort&compress}

\journal{Journal of Computational Physics}

\begin{document}

\begin{frontmatter}

\title{Performance potential for simulating spin models on GPU}

\author{Martin Weigel}
\ead{weigel@uni-mainz.de}
\ead[url]{http://www.cond-mat.physik.uni-mainz.de/~weigel}

\address{Institut f\"ur Physik, Johannes Gutenberg-Universit\"at Mainz,
  Staudinger Weg 7, D-55099 Mainz, Germany}

\address{Applied Mathematics Research Centre, Coventry University, Coventry, CV1~5FB, United Kingdom}

\begin{abstract}
  Graphics processing units (GPUs) are recently being used to an increasing degree
  for general computational purposes. This development is motivated by their
  theoretical peak performance, which significantly exceeds that of broadly available
  CPUs. For practical purposes, however, it is far from clear how much of this
  theoretical performance can be realized in actual scientific applications. As is
  discussed here for the case of studying classical spin models of statistical
  mechanics by Monte Carlo simulations, only an explicit tailoring of the involved
  algorithms to the specific architecture under consideration allows to harvest the
  computational power of GPU systems. A number of examples, ranging from Metropolis
  simulations of ferromagnetic Ising models, over continuous Heisenberg and
  disordered spin-glass systems to parallel-tempering simulations are
  discussed. Significant speed-ups by factors of up to $1000$ compared to serial
  CPU code as well as previous GPU implementations are observed.
\end{abstract}


\end{frontmatter}

\section{Introduction}

Numerical techniques and, in particular, computer simulations are by now firmly
established as a third pillar of the scientific method, complementing the pillars of
experiment (or observation) and mathematical theory, both of which were erected
already at the birth of modern science in the Scientific Revolution. While in the
early times \cite{metropolis:53a} simulation studies were not quite competitive with
analytical calculations in condensed matter physics and quantum field theory (usually
based on perturbative and variational approaches) nor were their outcomes adequate
for comparison with experimental results (usually due to limited length or time
scales), this situation has dramatically changed over the past decades. This very
successful race to catch up was fueled by a combination of two factors: a continuous,
sometimes revolutionary refinement of the numerical toolbox, for instance through the
invention of cluster algorithms \cite{kandel:91a}, reweighting \cite{ferrenberg:89a}
or generalized-ensemble techniques \cite{berg:92b,hukushima:96a} in the field of
Monte Carlo simulations, and the impressive increase in generally available
computational power, which has followed an exponential form known as Moore's law for
the past forty years. At any time, however, there has been no shortage of fascinating
scientific problems whose computational study required resources at or beyond the
cutting edge of the available technology. This has led scientists to regularly use
the latest commodity hardware as soon as it became available, but has also motivated
the design and construction of a number of special purpose machines such as, e.g.,
the cluster processor \cite{bloete:99a} and Janus \cite{belleti:09} for the
simulation of spin systems or a series of initiatives for QCD calculations with its
recent addition of the QPACE architecture \cite{goldrian:08}.

It has been true for the last couple of generations of graphics processing units
(GPUs) that their theoretical peak performance, in particular for single-precision
floating point operations (FLOP/s), significantly exceeds that of the corresponding
x86 based CPUs available at the same time (as of this writing up to around 100
GFLOP/s in CPUs vs.\ up to 5 TFLOP/s in GPUs). It is therefore natural that
scientists and, increasingly, also programmers of commercial applications other than
computer games, have recently started to investigate the possible value of GPU based
computing for their purposes; for scientific applications see, e.g.,
Refs.~\cite{tomov:05,meel:07,preis:09,bernaschi:10}. Apart from their mere peak
performance, systems based on GPUs or other highly parallel architectures such as the
Cell processor \cite{goldrian:08} might outperform current CPU based systems also in
terms of their efficiency, i.e., in terms of FLOP/s per Watt or FLOP/s per Euro and
thus might also contribute to the advancement of ``green computing''
\footnote{Incidentally, the Cell based QPACE system is ranked number one in the
  Green500 list of November 2009 \cite{green500}.}. The low prices and convenient
over-the-counter availability of GPU devices clearly make for advantages as compared
to custom-built special-purpose machines, for which many man-months or years need to
be invested for the design, realization and programming of devices. The relative
increase in peak performance of GPUs versus CPUs is achieved at the expense of
flexibility, however: while today's CPUs will perform pretty well on most of a large
variety of computer codes and, in particular, in a multi-tasking environment where
on-the-fly optimizations such as branch prediction are essential, GPUs are optimized
for the highly vectorized and parallelized floating-point computations typical in
computer graphics applications. A one-to-one translation of a typical code written
for CPUs to GPUs will therefore, most often, {\em not\/} run faster and, instead,
algorithms and parallelization and vectorization schemes must be chosen very
carefully to tailor for the GPU architecture in order to harvest any performance
increases. The efficiency of such calculations in terms of FLOP/s per {\em human\/}
time crucially depends on the availability of easily accessible programming
environments for the devices at hand. While in view of a lack of such supporting
schemes early attempts of general purpose GPU (GPGPU) calculations still needed to
encapsulate the computational entities of interest in OpenGL primitives
\cite{tomov:05}, the situation changed dramatically with the advent of
device-specific intermediate GPGPU language extensions such as ATI Stream and NVIDIA
CUDA \cite{cuda}. In the future, one hopes to become independent of specific devices
with general parallel programming environments such as OpenCL \cite{opencl}.

Classical spin systems have turned out to be extremely versatile models for a host
of phenomena in statistical, condensed matter and high-energy physics, with
applications ranging from critical phenomena \cite{pelissetto:02} over surface
physics \cite{degennes:85} to QCD \cite{svetitsky:82}. A rather significant effort,
therefore, has been invested over the years into optimized implementations of Monte
Carlo simulations of spin models. They hence appear to be particularly suited for an
attempt to fathom the potential gain from switching to the GPU
architecture. Additionally, there are a plethora of questions relating to classical
spin models which, decades of research notwithstanding, are still awaiting a
combination of an increase in available computational resources, better algorithms
and clever techniques of data analysis to find generally satisfactory answers. This
applies, in particular, to disordered systems such as random-field models and spin
glasses \cite{young:book}. As will be discussed below, due to their short-ranged
interactions and the typically simple lattice geometries such models appear to be
near ideal problems for GPU calculations. This applies, in particular, to the
inherently local single spin-flip algorithms discussed here. For studying the
critical behavior of models without disorder, cluster algorithms will outperform any
optimized implementation of a local spin-flip dynamics already for intermediate
lattice sizes; the possibilities for porting such approaches to GPU will be discussed
elsewhere \cite{weigel:10b}. For disordered systems, on the other hand, efficient
cluster algorithms are (with few exceptions \cite{houdayer:01}) not known. For them,
instead, local spin-flip simulations combined with parallel tempering
\cite{hukushima:96a} moves are the state of the art.

When comparing GPU and CPU performance for the case of general purpose computational
applications, it has become customary to benchmark different implementations in terms
of the relative speed-up (or slow-down) of the GPU code versus the CPU implementation
\cite{cuda,preis:09,dickson:10}. While such numbers make for good selling points for
the company producing the ``better'' type of computational device, it should be clear
that speed-ups, being a relative measure, will vary to a large degree depending on
how much effort is invested in the optimization of the codes for the different
architectures. To avoid this problem, the main focus is put here on the {\em
  absolute\/} performance of different implementations, measured for the present
application mostly in the wall-clock time for performing a single spin flip, citing
speed-up factors only as an additional illustration of the relative performance. If
relative measures of performance are given, the question arises of what type of CPU
code to compare to, in particular, since with the advent of multi-core processors and
vector extensions such as SSE, CPUs also offer a significant potential for
optimizations. I decided here to use serial CPU code, reasonably optimized on the
level of a high-level programming language and the use of good compilers, as I feel
that this comes closest to what is most often being used in actual simulation
studies. As regards the possible effects of further CPU optimizations, see also
Ref.~\cite{dickson:10} which, however, unfortunately does not cite any measures of
absolute performance. Simulations of the ferromagnetic Ising model using
implementations on GPU have been discussed before
\cite{tomov:05,preis:09,block:10,levy:10,hawick:10a}. Compared to these
implementations, the current approach with the double checkerboard decomposition and
multi-hit updates to be discussed below offers significant advantages. Other
applications, such as the simulation of ferromagnetic, short-range Heisenberg models,
the simulation of Ising spin glasses with asynchronous multi-spin coding or parallel
tempering for lattice spin systems are considered here for the first time. For some
very recent discussions of further spin models see also
Refs.~\cite{tapia:11,campos:11}.

The rest of this article is organized as follows. In Sec.~\ref{sec:architecture}, I
give some necessary background on the architecture of the NVIDIA GPUs used in this
study and its consequences for algorithm design. Section \ref{sec:ising} discusses,
in some detail, the chosen implementation of a single-spin flip Metropolis simulation
of the two-dimensional (2D) Ising model and its performance as well as a
generalization to the three-dimensional (3D) model. Section \ref{sec:heisen} is
devoted to generalizations of these considerations to continuous-spin systems,
exemplified in the 2D Heisenberg model.  In Secs.~\ref{sec:spinglass} and
\ref{sec:tempering}, applications to simulations of spin-glass models and the use of
parallel-tempering updates are discussed. Finally, Sec.~\ref{sec:concl} contains my
conclusions.

\section{The NVIDIA architecture\label{sec:architecture}}

As indicated above, there are significant differences in the general design of CPU
and GPU units \cite{owens:08}. CPUs have been optimized over the past decades to be
``smart'' devices for the rather unpredictable computational tasks encountered in
general-purpose computing.  Current Intel CPUs feature about 800 million transistors
on one die. A rather small fraction of them is used for the ALUs (arithmetic logic
units) doing actual computations, whereas most of the available transistors are
devoted to flow control (such as out-of-order execution and branch prediction) and
cache memory. This structure is very successful in increasing the performance of {\em
  single}-threaded applications, which still make up the vast majority of general
computer codes. In contrast, about 80~\% of the 1.4 billion transistors on a GPU die
of the NVIDIA GT200 series (now superseded by the Fermi architecture) are ALUs. GPUs
do not try to be ``clever'', but they are extremely efficient at doing the same
computational steps on different bits of a large data-set in parallel. This is what
makes them interesting for applications in scientific computing.

\begin{figure}[tb]
  \centering
  \includegraphics[width=0.9\textwidth]{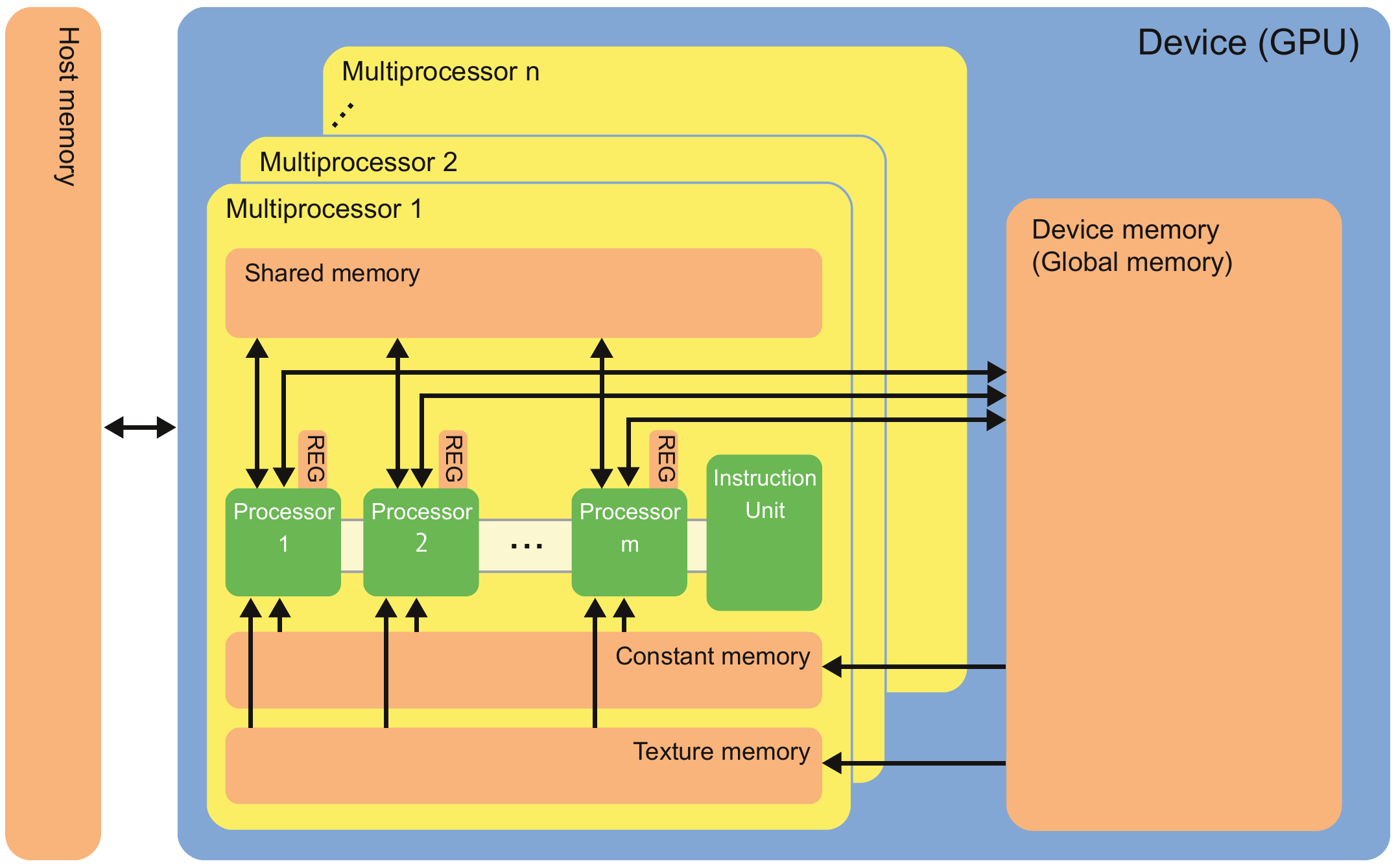}
  \caption{Schematic representation of the architecture of current GPUs. The
    terminology is borrowed from that used for NVIDIA boards.}
  \label{fig:hardware}
\end{figure}

Figure \ref{fig:hardware} shows a schematic representation of the general
architecture of a current GPU. The chip contains a number of multiprocessors each
composed of a number of parallel processing units. The systems based on the GT200
architecture used in this study feature 30 multiprocessors of 8 processors each
(versus 15 multiprocessors \`a 32 cores for the GTX 480 Fermi card). The systems come
with a hierarchy of memory layers with different characteristics. Each processor is
equipped with a number of registers which can be accessed with essentially no
latency. The processors in a multiprocessor unit have read/write access to a small
amount of shared memory ($16$ KB in the GT200 architecture and $48$ KB for Fermi
cards) which is on-chip and can be accessed with latencies around a hundred times
smaller than those for global memory. The large device or global memory (up to 6 GB
in current Tesla systems) is connected with read/write access to all
multiprocessors. Latency for global memory accesses is of the order of 400--600 clock
cycles (as compared to, e.g., one clock cycle for a multiply or add operation). The
additional constant and texture memory areas are cached and designed for read-only
access from the processing units, in which case they operate very fast with small
latencies. The memory of the host computer cannot be accessed directly from within
calculations on the GPU, such that all relevant data need to be copied into and out
of the GPU device before and after the calculation, respectively. Since the
processing units of each multiprocessor are designed to perform identical
calculations on different parts of a data set, flow control for this SIMD (single
instruction multiple data) type of parallel computations is rather simple. It is
clear that this type of architecture is near ideal for the type of calculations
typical for computer graphics, namely rendering a large number of triangles in a 3D
scene or the large number of pixels in a 2D projection in parallel.

\begin{figure}[tb]
  \centering
  \includegraphics[width=0.9\textwidth]{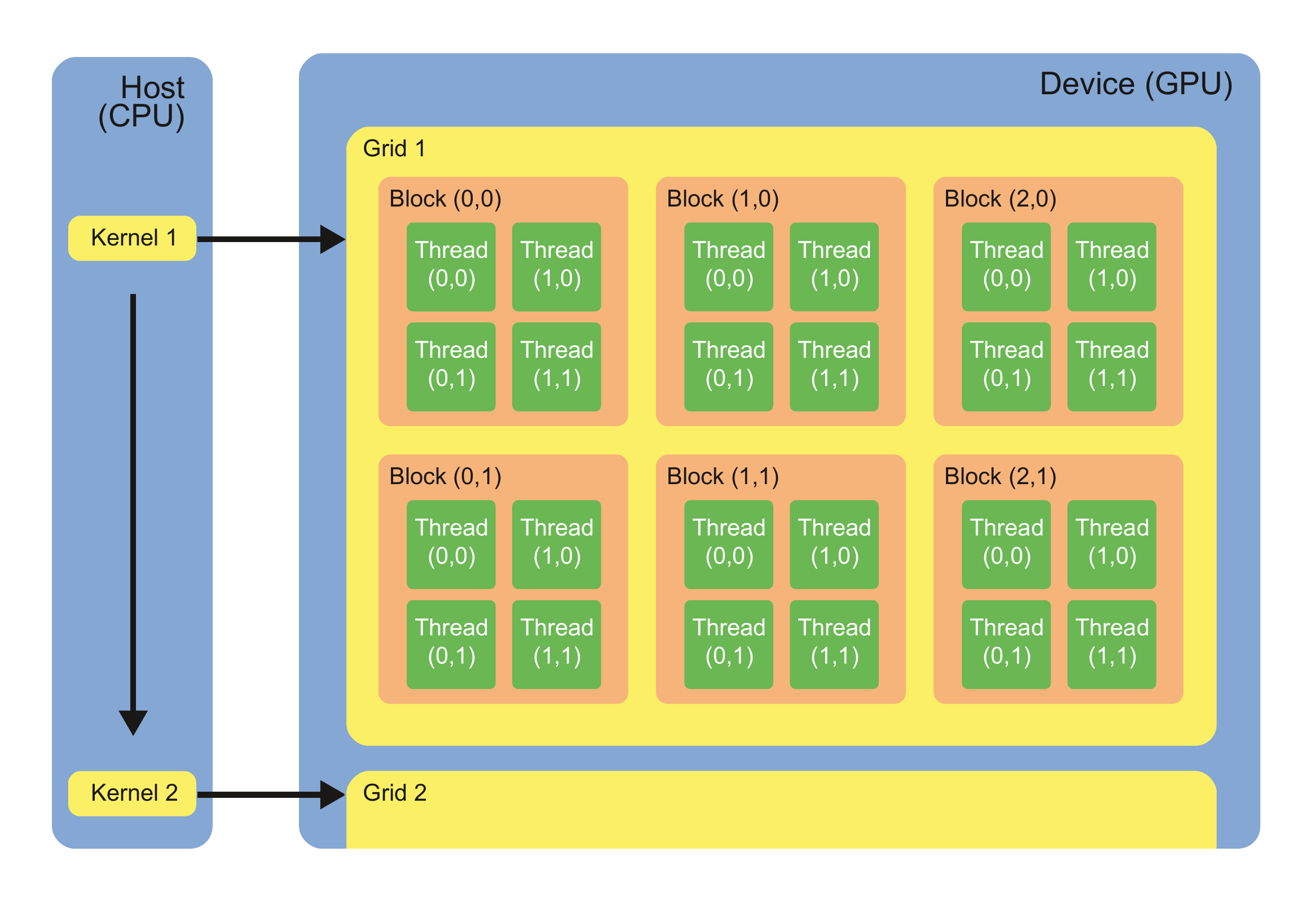}
  \caption{Parallel execution of a GPU program (``kernel'') in a grid of thread
    blocks. Threads within a block work synchronously on the same data set. Different
    blocks are scheduled for execution independent of each other.}
  \label{fig:gridblock}
\end{figure}

The organization of processing units and memory outlined in Fig.~\ref{fig:hardware}
translates into a combination of two types of parallelism: the processing units
inside of each multiprocessor work synchronously on the same data set
(vectorization), whereas different multiprocessors work truly independent of each
other (parallelization). The corresponding programming model implemented in the CUDA
framework \cite{cuda} is outlined schematically in Fig.~\ref{fig:gridblock}:
computations on GPU are encapsulated in functions (called ``kernels'') which are
compiled to the GPU instruction set and downloaded to the device. They are executed
in a two-level hierarchic set of parallel instances (``execution configuration'')
called a ``grid'' of thread ``blocks''. Each block can be thought of as being
executed on a single multiprocessor unit. Its threads (up to 512 for the GT200
architecture and 1024 for Fermi cards) access the same bank of shared memory
concurrently. Ideally, each thread should execute exactly the same instructions, that
is, branching points in the code should be reduced to a minimum. The blocks of a grid
(up to $65536\times 65536$) are scheduled independently of each other and can only
communicate via global memory accesses. The threads within a block can make use of
cheap synchronization barriers and communicate via the use of shared (or global)
memory, avoiding race conditions via atomic operations implemented directly in
hardware. On the contrary, the independent blocks of a grid cannot effectively
communicate within a single kernel call. If synchronization between blocks is
required, consecutive calls of the same kernel are required, since termination of a
kernel call enforces all block computations and pending memory writes to complete.

Since the latency of global memory accesses is huge as compared to the expense of
elementary arithmetic operations, many computational tasks on GPU will be limited by
memory accesses, i.e., the bandwidth of the memory subsystem. It is therefore crucial
for achieving good performance to (a) increase the number of arithmetic operations
per memory access, and (b) optimize memory accesses by using shared memory and
choosing appropriate memory access patterns. The latter requirement in particular
includes the adherence to the specific {\em alignment\/} conditions for the different
types of memory and clever use of the {\em coalescence} phenomenon, which means that
accesses to adjacent memory locations issued by the different threads in a block
under certain conditions can be merged into a single request, thus greatly improving
memory bandwidth. Due to most typical computations being limited by memory accesses,
it is important to use an execution configuration with a total number of threads
(typically several $1000$) much larger than the total number of processing units
($240$ for the Tesla C1060 and $480$ for the GTX 480). If a thread block issues an
access, e.g., to global memory, the GPU's scheduler will suspend it for the number of
cycles it takes to complete the memory accesses and, instead, execute another block
of threads which has already finished reading or writing its data. The good
performance of these devices thus rests on the idea of latency hiding and transparent
scalability through flexible thread scheduling \cite{kirk:10}.

\section{Metropolis simulations of the ferromagnetic Ising model\label{sec:ising}}

The layout of the GPU architecture outlined above implies guidelines for the
efficient implementation of computer simulation codes. Along these lines, a code for
single-spin flip Metropolis simulations of the ferromagnetic Ising model is
developed.

\subsection{General considerations}

We consider a ferromagnetic, nearest-neighbor, zero-field Ising model with
Hamiltonian
\begin{equation}
  \label{eq:ising_hamiltonian}
  {\cal H} = -J\sum_{\l i,j\r}s_i s_j,\;\;\;s_i = \pm 1
\end{equation}
on square and simple cubic lattices of edge length $L$, using periodic boundary
conditions. In the single-spin flip Metropolis simulation scheme for this model each
step consists of randomly selecting a spin $i\in\{1,\ldots,N=L^d\}$ and proposing a
spin flip $s_i \mapsto -s_i$, which is accepted according to the Metropolis criterion
\cite{metropolis:53a}
\begin{equation}
  \label{eq:metropolis}
  p_\mathrm{acc}(s_i \mapsto -s_i) = \min\left[1,\,e^{-\beta\Delta E}\right],
\end{equation}
where $\Delta E/J = 2 s_i \sum_{j\;\mathrm{nn}\;i} s_j$ corresponds to the energy
change incurred by the flip and $\beta = 1/k_B T$ denotes the inverse temperature. It
is straightforward to show that this update is ergodic, i.e., all states of the
system can be reached in a finite number of spin-flip attempts with non-zero
probability, and satisfies the detailed balance condition ensuring convergence to the
equilibrium Boltzmann distribution. In practice, one usually walks through the
lattice in a sequential fashion instead of picking spins at random, which requires
less pseudo-random numbers and, generically, leads to somewhat faster
convergence. This updating scheme, in fact, violates detailed balance, but is
consistent with the global balance condition which is sufficient to ensure
convergence \cite{berg:04}. This point is of some importance for the GPU
implementation developed here, and will be further discussed below in Section
\ref{sec:balance}.

\subsection{Double checkerboard decomposition\label{sec:checker}}

To leverage the effect of the massively parallel architecture of GPU devices for
simulations of spin models, in most cases domain decompositions where the system is
divided into a large number of largely independent sub-units are the only approach
with satisfactory scaling properties as the number of processors or the size of the
system is increased. For the case of lattice systems, the simplest scheme amounts to
a coloring of lattice sites with the minimally feasible number of colors. Here, I
focus on bipartite graphs such as the square and (hyper-)cubic lattices where two
colors suffice, resulting in a (generalized) checkerboard decomposition
\cite{heermann:90}. Generalizations to other cases are trivial.

\begin{figure}[tb]
  \centering
  \includegraphics[width=0.6\textwidth]{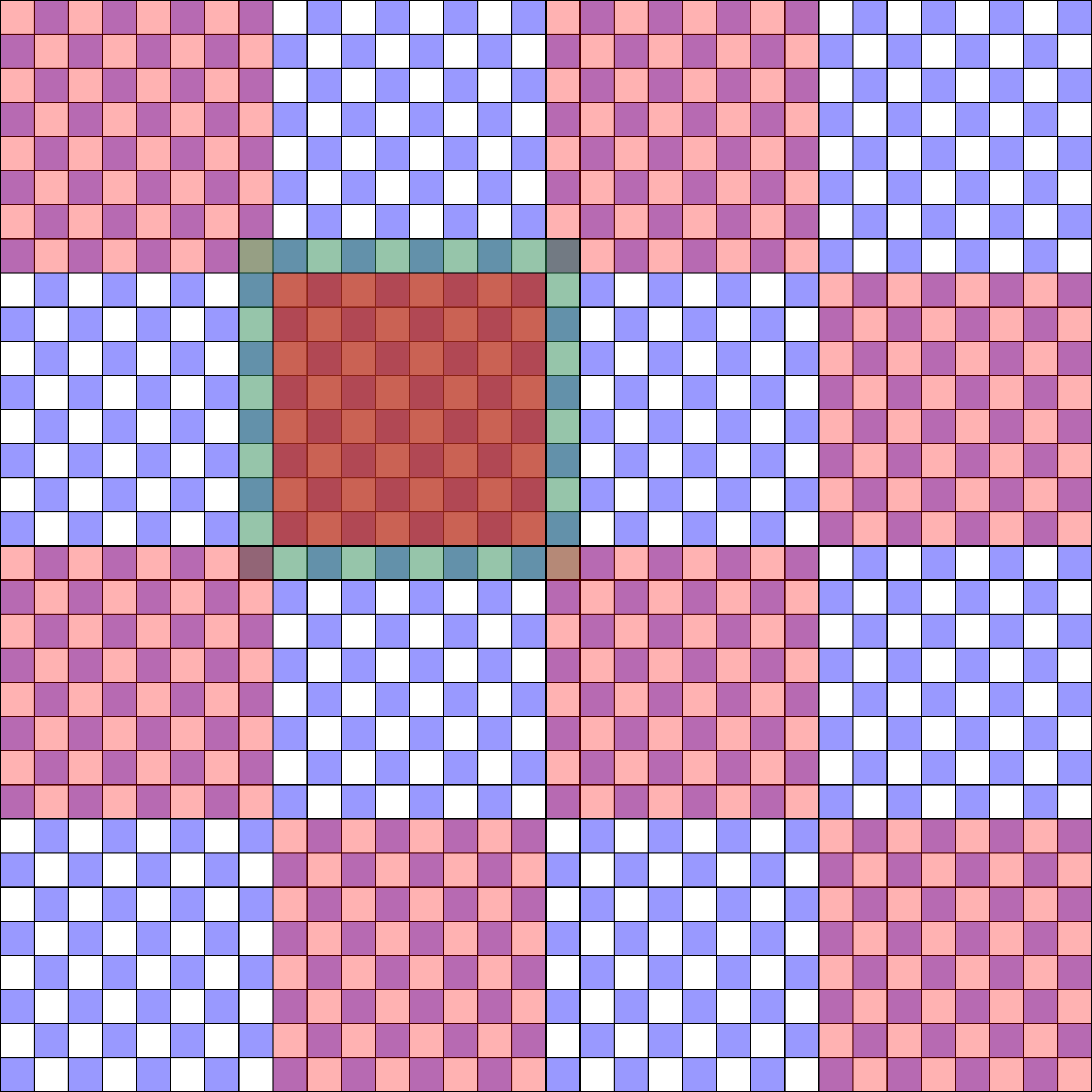}
  \caption{Double checkerboard decomposition of a square lattice of edge length $L =
    32$ for performing a single spin-flip Metropolis simulation of the Ising model on
    GPU. Each of the $B\times B = 4\times 4$ big tiles is assigned as a thread block
    to a multiprocessor, whose individual processors work on one of the two
    sub-lattices of all $T\times T = 8\times 8$ sites of the tile in parallel.}
  \label{fig:checker}
\end{figure}

In such a scheme, each site of one sub-lattice can be updated independently of all
others (assuming nearest-neighbor interactions only), such that all of them can be
treated in parallel followed by an analogous procedure for the second
sub-lattice. For an implementation on GPU, the checkerboard decomposition needs to be
mapped onto the structure of grid blocks and threads. Although, conceptually, a
single thread block suffices to update one sub-lattice in parallel, the limitation to
512 threads per block for the GT200 architecture (resp.\ 1024 threads for Fermi)
enforces an additional block structure for systems of more than 1024 resp.\ 2048
spins. For the square lattice, in Ref.~\cite{preis:09} stripes with a height of two
lattice spacings were assigned to independent thread blocks while performing all spin
updates in global memory. This has several disadvantages, however: (a) shared memory
with its much lower latency is not used at all, (b) determining the indices of the
four neighboring spins of each site requires costly conditionals due to the periodic
boundary conditions, and (c) the system size is limited to $L\le 1024$ for GT200
($L\le 2048$ for Fermi). Here, instead, I suggest a more versatile and efficient
approach leveraging the inherent power of the memory hierarchy intrinsic to GPUs. To
this end a twofold hierarchy of checkerboard decompositions is used. Figure
\ref{fig:checker} illustrates this for the case of a square lattice: on the first
level, the lattice is divided into $B\times B$ big tiles in a checkerboard
fashion. These are then updated by individual thread blocks, where the individual
threads exploit a second-level checkerboard decomposition of the $T\times T$ lattice
sites inside each tile. The size of big tiles is thereby chosen such that the spin
configuration on the tile (plus a boundary layer of spins) fits into shared memory,
such that the spin updates can then be performed entirely inside of this much faster
memory area. On the block level, through the checkerboard arrangement all tiles of
one (``even'') sub-lattice can be updated concurrently before updating the other
(``odd'') half of tiles. Inside of each block, again all sites of the finer
sub-lattices are independent of each other and are therefore updated concurrently by
the threads of a thread block.

In summary, the updating procedure looks as follows:
\begin{enumerate}
\item The updating kernel is launched with $B^2/2$ thread blocks assigned to treat
  the {\em even\/} tiles of the coarse checkerboard.
\item The $T^2/2$ threads of each thread block cooperatively load the spin
  configuration of their tile plus a boundary layer into shared memory.
\item The threads of each block perform a Metropolis update of each {\em even\/}
  lattice site in their tile in parallel.
\item All threads of a block wait for the others to finish at a barrier
  synchronization point.
\item The threads of each block perform a Metropolis update of each {\em odd\/}
  lattice site in their tile in parallel.
\item The threads of each block are again synchronized.
\item A second kernel is launched working on the $B^2/2$ {\em odd\/} tiles of the
coarse checkerboard in the same fashion as for the even tiles.
\end{enumerate}
The cooperative loading of each tile into shared memory is organized in a fashion to
ensure {\em coalescence\/} of global memory accesses \cite{kirk:10}, i.e., subsequent
threads in each block access consecutive global memory locations wherever
possible. While it turns out to be beneficial to load tiles into shared memory
already for a single update per spin before treating the second coarse sub-lattice
due to an avoidance of bank conflicts in global memory as well as the suppressed need
to check for periodic wrap-arounds in determining lattice neighbors, the ratio of
arithmetic calculations to global memory accesses is still not very favorable. This
changes, however, if a multi-hit technique is applied, where the spins on each of the
two coarse sub-lattices are updated several times before returning to the other half
of the tiles. As is discussed below in Sec.~\ref{sec:balance}, this works well, in
general, and only close to criticality somewhat increased autocorrelation times are
incurred.

\subsection{Random number generation\label{sec:rng}}

By design, Monte Carlo simulations of spin systems require a large amount of
pseudo-random numbers. Depending on implementation details, for a system as simple as
the Ising model, the time required for generating random numbers can dominate the
total computational cost. Hence, the efficient implementation of random-number
generators (RNGs) in a massively parallel environment is a crucial requirement for
the efficient use of the GPU architecture. Speed is not everything, however, and it
is well known that the statistical deficiencies that no {\em pseudo\/} RNG can
entirely avoid can have rather dramatic consequences in terms of highly significant
deviations of simulation results from the correct answers
\cite{ferrenberg:92,shchur:97}. Hence, some effort must be invested in the choice and
implementation of an appropriate RNG.

From the highly parallel setup described above, it is clear that each of the $N =
L^2/4$ concurrent threads must be able to generate its stream of random numbers to a
large degree independently of all others, since the bottleneck in any configuration
with a centralized random-number production would severely impede scaling of the code
to a large number of processing units. As each of these (sub-)sequences of random
numbers are used in the same simulation, one needs to ensure that (a) either each
thread generates different members of the same global sequence of numbers or (b) the
sequences generated by different threads are at least statistically uncorrelated. The
simplest choice of RNG is given by the linear congruential generator (LCG) of the
form
\begin{equation}
  \label{eq:lcg}
  x_{n+1} = (ax_n+c)\,\mathrm{mod}\,m
\end{equation}
with $m = 2^{32}$. The authors of Ref.~\cite{preis:09} used $a = 1664525$ and $c =
1013904223$, originally suggested in Ref.~\cite{numrec}. The period of this generator
is small with $p = 2^{32} \approx 4\times 10^9$ \cite{knuth:vol2}. In a simulation of
a $4096\times 4096$ Ising system, for instance, this period is exhausted already
after 256 sweeps. On theoretical grounds, it is argued that one actually should not
use more than $\sqrt{p}$ numbers out of such a sequence \cite{knuth:vol2,gentle:03},
which would render the sequence of available numbers very short indeed. Additionally,
simple LCGs are known to exhibit strong correlations which can be revealed by
plotting $k$-tuples of successive (normalized) numbers as points in $\mathbb{R}^k$,
where it is found that, already for rather small $k$, the points are confined to a
sequence of hyperplanes instead of being uniformly distributed. The choice $m =
2^{32}$ has the advantage that the whole calculation can be implemented entirely in
32-bit integer arithmetic since on most modern architectures (including GPUs) integer
overflows are equivalent to taking a modulo operation. For such power of two moduli
$m$, however, the period of the less significant bits is even shorter than that of
the more significant bits, such that the period of the $k$th least significant bit is
only $2^k$. An advantage for the parallel calculations performed here is that one can
easily skip ahead in the sequence, observing that
\begin{equation}
  x_{n+t} = (a_t x_n+c_t)\,\mathrm{mod}\,m,
  \label{eq:lcg_skipping}
\end{equation}
where
\begin{equation}
  a_t = a^t\,\mathrm{mod}\,m,\;\;\;
  b_t = \sum_{i=1}^t a^i c \,\mathrm{mod}\,m.
\end{equation}
Therefore, choosing $t$ equal to the number $N=L^2/4$ of threads, all threads can
generate numbers out of the same global sequence (\ref{eq:lcg}) concurrently. An
alternative setup, that cannot guarantee the desired independence of the sequences
associated to individual RNG instances, however, starts from randomized initial seeds
for each generator, without using any skip-ahead \cite{preis:09}. To improve on the
drawback of a short period, one might consider moving to a generator with larger $m$,
for instance $m= 2^{64}$, where the modulo operation again can be implemented by
overflows, this time for $64$-bit integers, a data type which is also available in
the CUDA framework. As multiplier I chose $a = 2862933555777941757$ with provably
relatively good properties \cite{lecuyer:99}, where an odd offset, here $c =
1442695040888963407$, needs to be chosen to reach the maximal period of $p = 2^{64}
\approx 2\times 10^{19}$. As for the $32$-bit case, this generator can be used in the
parallel framework yielding numbers from a common sequence via skip-ahead, or as
independent generators with initial randomized seeds, leading to overlapping
sub-sequences.

\begin{figure}[tb]
  \centering
  \includegraphics[keepaspectratio=true,scale=0.75,trim=45 48 75 78]{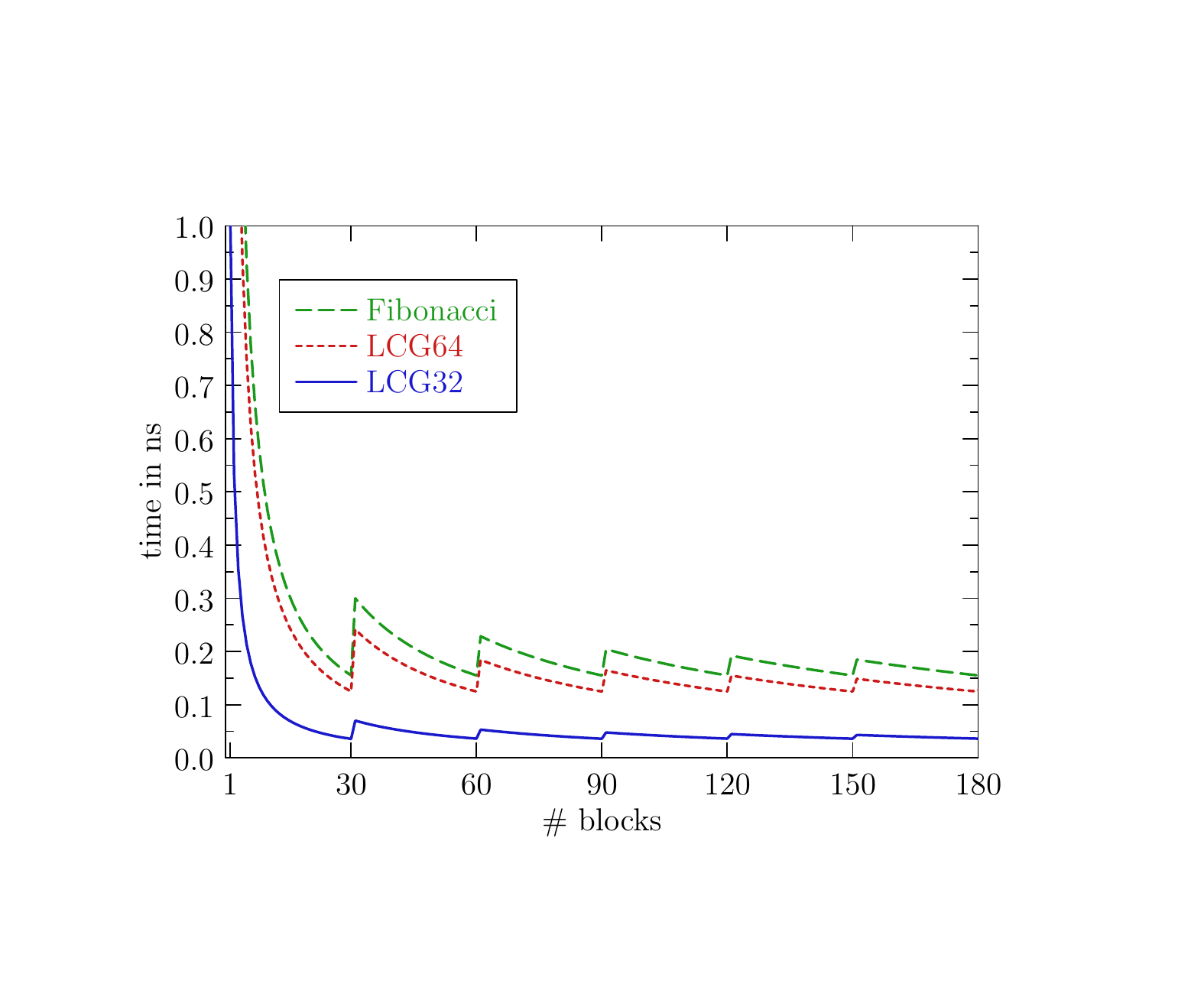}
  \caption{Computer time per random number for running different implementations of
    pseudo RNGs on a Tesla C1060 GPU as a function of the number of grid blocks
    employed. The ``Fibonacci'' generator corresponds to the recursion
    \eqref{eq:lagged_fibonacci2} with $\alpha = \beta = 1$ and $r = 521$, $s = 353$,
    while ``LCG32'' and ``LCG64'' correspond to the recursion \eqref{eq:lcg} with $m
    = 2^{32}$ and $m = 2^{64}$, respectively. All thread blocks use $512$ threads.}
  \label{fig:rng}
\end{figure}

For high-precision calculations, one normally would not want to rely on a simple
LCG. A significant number of generators with much better statistical properties has
been developed in the past, for a review see Ref.~\cite{gentle:03}. For our purposes,
however, most of them have severe drawbacks in that they are either quite slow (such
as, for instance, generators that combine a number of elementary RNGs or, instead,
drop a certain fraction of random numbers generated as in RANLUX \cite{luescher:94}),
that they use a large state to operate on (for instance $624$ 32-bit words for the
popular Mersenne twister \cite{matsumoto:98}), or that it is rather hard to ensure
that sequences generated independently are to a reasonable degree uncorrelated (as
for most high-quality generators). While a state larger than a few words is usually
not a problem for serial implementations, in the framework of GPU computing where
fast local (i.e., shared) memory is very limited, it is impossible to use several
hundred words per thread of a thread block only for random number
generation\footnote{Note that the ``official'' CUDA implementation of Mersenne
  twister \cite{cuda} uses data structures entirely in global memory.}. A reasonable
compromise could be given by generators of the (generalized) lagged Fibonacci type
with recursion
\begin{equation}
  \label{eq:lagged_fibonacci}
  x_{n} = a x_{n-r} \oplus b x_{n-s}\,\mathrm{mod}\,m,
\end{equation}
which operate on a buffer of length $r+s$ with $s<r$ and have good properties for
sufficiently large lags $r$ and $s$ and choosing $\oplus = \pm$ \cite{brent:92}. If
$m = 2^w$, the maximal period is $p = 2^{w-1}(2^r-1)$. For an implementation in
single precision arithmetic, i.e., $w = 32$, and $r = 1279$ (see below), this results
in a rather large period $p \approx 2\times 10^{394}$. The recursion
(\ref{eq:lagged_fibonacci}) can be implemented directly in floating-point arithmetic
yielding uniform random numbers $u_i\in[0,1]$ via
\begin{equation}
  \label{eq:lagged_fibonacci2}
  u_{n} = \alpha u_{n-r} + \beta u_{n-s}\,\mathrm{mod}\,1.
\end{equation}
For good quality one needs $r \gtrsim 100$, leading to relatively large storage
requirements, but here the generation of $s$ random numbers can be vectorized by the
threads of a thread block. Therefore, the required ring buffer of length $r+s$ words
is shared by the $T^2/2$ threads of a block, and hence the storage requirement in
only $2(r+s)/T^2$ words per thread. If one chooses $s$ only slightly larger than the
number of threads per block, only a few words per thread are consumed. A number of
good choices for the lags $r$ and $s$ are collected in Ref.~\cite{brent:92}. The best
performance is achieved for unity multipliers $\alpha = \beta = 1$, at only
moderately reduced quality of the resulting random numbers. In this setup, different
thread blocks use different sequences. Prescriptions for ensuring disjunct sequences
are described in Ref.~\cite{brent:92}. In view of the astronomic period, however, it
appears safe to just seed the ring buffers of the generators for different blocks
with an independent RNG.

The LCG generator \eqref{eq:lcg} with $m=2^{32}$ and $m=2^{64}$ as well as the
generalized Fibonacci generator \eqref{eq:lagged_fibonacci2} were implemented and
benchmarked on a Tesla C1060 GPU. The LCGs have been realized with $32$- and $64$-bit
integers, respectively, using the multipliers listed above. For the Fibonacci
generator, I chose $\alpha = \beta = 1$ and lags $r = 521$ and $s = 353$ (suitable
for simulations with tile size up to $T = 16$). The generators were test with a
variable number of blocks of $512$ threads each (where $512$ is the maximum number of
threads allowed on the C1060 device). All generator state information is kept in
shared memory. The results are displayed in Fig.~\ref{fig:rng}. The characteristic
zig-zag pattern results from the commensurability (or not) of the number of blocks
$b$ with the number of $30$ multiprocessors on the device, such that the performance
is best if $b = 30\ell$, $\ell=1$, $2$, $\ldots$ and worst if $b = 30\ell+1$, since
in the latter case $29$ of the multiprocessors are idle once the last batch of blocks
has been scheduled for execution. The $32$-bit LCG is found to take at best $0.036$
ns per random number, whereas the $64$-bit LCG and the lagged Fibonacci generator are
about $3.5$ and $4.3$ times slower, respectively.

\begin{table}[tb]
  \centering
  \caption{Internal energy $e$ per spin and specific heat $C_V$ for a $1024\times 1024$
    Ising model with periodic boundary conditions at $\beta = 0.4$
    from simulations on CPU and on GPU
    using different random number generators. ``LCG32'' and ``LCG64'' correspond to
    the recursion (\ref{eq:lcg}) with $m = 2^{32}$ and $m = 2^{64}$, respectively,
    while ``Fibonacci'' denotes the generator (\ref{eq:lagged_fibonacci2}) with
    $r = 521$, $s = 353$ and $r=1279$, $s=861$, respectively.
    For the ``random'' versions of LCGs on GPU, the generators
    for each thread are seeded randomly, while in the other cases all threads
    sample from the same sequence using skipping according to
    Eq.~(\ref{eq:lcg_skipping}). $\Delta_\mathrm{rel}$ denotes the deviation from
    the exact result relative to the estimated standard deviation.}
  \vspace*{2ex}
  \begin{tabular}{lt{9}t{2}t{7}.} \hline
    \multicolumn{1}{c}{method} & \multicolumn{1}{c}{$e$} &
    \multicolumn{1}{c}{$\Delta_\mathrm{rel}$} &
    \multicolumn{1}{c}{$C_V$} &
    \multicolumn{1}{c}{$\Delta_\mathrm{rel}$} \\ \hline
    exact  & 1.106079207   & 0 & 0.8616983594 & 0 \\ \hline
    \multicolumn{5}{c}{sequential update (CPU)} \\ \hline
    LCG32              & 1.1060788(15)      & -0.26    & 0.83286(45) & -63.45 \\
    LCG64              & 1.1060801(17)      & 0.49     & 0.86102(60) & -1.14 \\
    Fibonacci, $r=512$ & 1.1060789(17)      & -0.18    & 0.86132(59) & -0.64 \\ \hline
    \multicolumn{5}{c}{checkerboard update (GPU)} \\ \hline
    LCG32              & 1.0944121(14)      & -8259.05 & 0.80316(48) & -121.05 \\
    LCG32, random      & 1.1060775(18)      & -0.97    & 0.86175(56) & 0.09 \\
    LCG64              & 1.1061058(19)      & 13.72    & 0.86179(67) & 0.14 \\
    LCG64, random      & 1.1060803(18)      & 0.62     & 0.86215(63) & 0.71 \\
    Fibonacci, $r=512$ & 1.1060890(15)      & 6.43     & 0.86099(66) & -1.09 \\
    Fibonacci, $r=1279$& 1.1060800(19)      & 0.40     & 0.86084(53) & -1.64 \\ \hline
  \end{tabular}
  \label{tab:rng}
\end{table}

The quality of the streams of random numbers generated in this way was tested with
the Metropolis update simulation of the 2D Ising model using the double checkerboard
update. The resulting estimates for the internal energy and specific heat of a system
of edge length $L = 1024$ and inverse temperature $\beta = 0.4$ from simulations
employing a number of different generators are collected in Tab.~\ref{tab:rng}. For
comparison, the exact results calculated according to Ref.~\cite{ferdinand:69a} are
shown as well. For the sequential updates on CPU, the dominant correlations and short
period of the $32$-bit LCG lead to highly significant deviations at the chosen rather
high level of statistics ($10^7$ updates per spin), which are, however, only visible
in the specific heat data. Already the $64$-bit LCG and likewise the Fibonacci
generator result in data completely consistent with the exact values at the given
level of statistical accuracy. The double checkerboard simulations on GPU, on the
other hand, appear to be much more sensitive to correlations in the RNGs. If all
threads sample from the same sequence of random numbers using skipping according to
Eq.~\eqref{eq:lcg_skipping}, the $32$-bit LCG produces astronomic and the $64$-bit
LCG still sizable deviations from the exact result, cf.\ the data collected in
Tab~\ref{tab:rng}. Somewhat surprisingly, no significant deviations are produced from
either LCG when seeding the LCG of each thread randomly and independently with
another RNG. In some sense, this setup appears to mimic the effects of a combined
generator \cite{numrec}: although all of the sequences traversed by individual
threads have significant overlap with each other, the random shifts between different
sequences are sufficient to cancel the correlation effects. The Fibonacci generators
of Eq.~\eqref{eq:lagged_fibonacci2}, on the other hand, do not suffer from such
problems, although deviations are again larger than for the sequential
implementation, and one needs to go to a lag as large as $r = 1279$ to get fully
satisfactory results for the system at hand. Note that the observed differences
between CPU and GPU results are entirely due to the different order of spin updates
in the sequential and checkerboard setups, and have nothing to do with shortcomings
of the GPU architecture. In particular, all the generators (and variants) implemented
here on GPU in connection with the corresponding simulation codes guarantee
completely reproducible results, i.e., the sequence of generated spin configurations
only depends on the initial seeding of the RNGs and not the order in which different
blocks or threads are executed.

For general practical applications, the use of the lagged Fibonacci generator appears
to be a reasonable compromise in that it is only rather insignificantly slower than
the $64$-bit LCG, but produces random numbers of substantially better quality. The
setup with randomly seeded independent LCGs per thread also yields satisfactory
results (and can be very fast), but in the absence of a deeper understanding of how
correlations are destroyed in this setup, it cannot be entirely recommended for
high-precision studies of simple, discrete models such as the Ising ferromagnet
considered here. For systems with continuous degrees of freedom or with additional
quenched disorder, however, such correlation effects are presumably less of a
problem.

\subsection{Balance and autocorrelations\label{sec:balance}}

It is worthwhile to note that the checkerboard update, like the sequential update of
lattice sites usually applied in a scalar computing context, does not satisfy
detailed balance (see, e.g., Ref.~\cite{berg:04}),
\begin{equation}
  \label{eq:detailed_balance}
  T(\{s_i'\}\rightarrow\{s_i\})p_\beta(\{s_i'\}) =
  T(\{s_i\}\rightarrow\{s_i'\})p_\beta(\{s_i\}), 
\end{equation}
on the level of single spin flips since the probability of selecting a spin twice for
updating in direct succession vanishes if the lattice sites are treated in a
deterministic order. The global balance condition,
\begin{equation}
  \label{eq:global_balance}
  \sum_{\{s_i'\}}T(\{s_i'\}\rightarrow\{s_i\})p_\beta(\{s_i'\}) = 
          \sum_{\{s_i'\}}T(\{s_i\}\rightarrow\{s_i'\})p_\beta(\{s_i\}) = p_\beta(\{s_i\}),
\end{equation}
on the other hand, which is sufficient to ensure convergence of the chain
\cite{behrends:02}, stays satisfied at all times. The same holds true if the
checkerboard update is applied in a multi-hit fashion with each coarse sub-lattice
being updated $k \ge 1$ times before moving on to the other sub-lattice. Although
detailed balance is not required for sampling the equilibrium distribution, it can be
recovered, if desired, e.g., to ensure that certain dynamical properties hold, on the
level of compound moves such as lattice sweeps. Denote, for instance, an update of
the even sub-lattice with ``A'' and an update of the odd sub-lattice with ``B'' and a
measurement time by ``M''. Then, a sequence of the form
$$
   AAAA(M)AAAABBBB(M)BBBBAAAA(M)AAAABBBB(M)BBBB\cdots
$$
satisfies detailed balance with respect to compound updates and is hence more
symmetric than the naive multi-hit implementation.

\begin{figure}[tb]
  \centering
  \includegraphics[keepaspectratio=true,scale=0.75,trim=45 48 75 78]{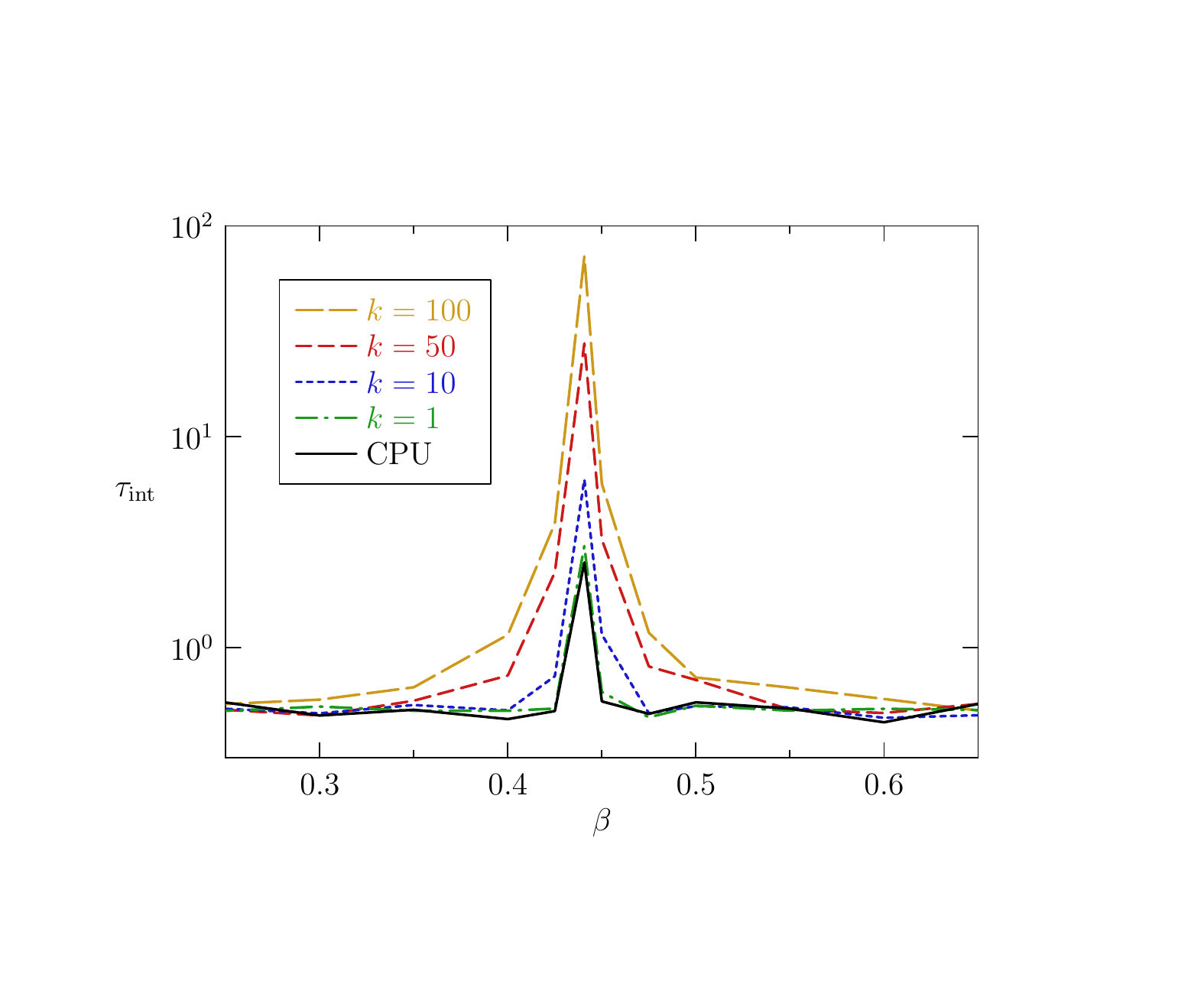}
  \caption{Integrated autocorrelation times of the internal energy for simulations of
    an $128\times 128$ Ising model as a function of inverse temperature $\beta$. The
    GPU calculations were performed with different numbers $k$ of multi-hit updates,
    but at a constant total number of updates.}
  \label{fig:autocorr}
\end{figure}

\begin{figure}[tb]
  \centering
  \includegraphics[keepaspectratio=true,scale=0.75,trim=45 48 75 78]{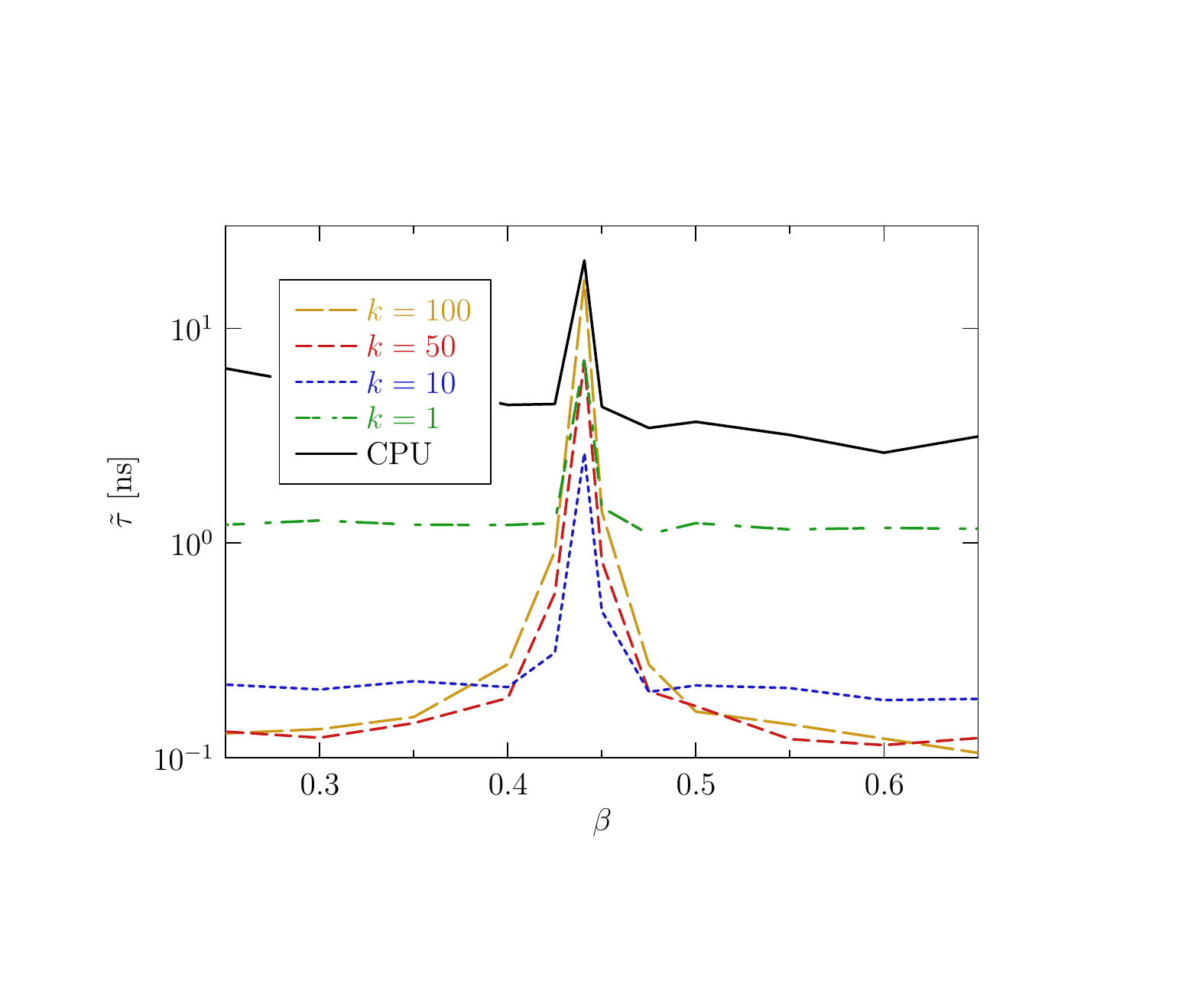}
  \caption{Re-scaled autocorrelation times $\tilde{\tau}_\mathrm{int} =
    \tau_\mathrm{int} t_\mathrm{flip}$ for the $128\times 128$ Ising system and
    double checkerboard updates on GPU for different values of the multi-hit
    parameter $k$. Minimal $\tilde{\tau}_\mathrm{int}$ corresponds to a minimum time
    of generating a statistically independent spin configuration.}
  \label{fig:autocorr2}
\end{figure}

While the multi-hit approach with $k>1$ is a perfectly valid updating scheme, it is
clear that very close to the critical point it will be somewhat slower than the $k=1$
version at decorrelating the system as soon as the correlation length $\xi$ exceeds
the size $T$ of the tiles. Assuming that $\xi \gtrsim T$, it takes
$$
\tau_\mathrm{tile} \sim T^2
$$
sweeps to decorrelate the configuration of the tile or, equivalently, for information
traveling diffusively to cross the tile. For $k \gg \tau_\mathrm{tile}$, one expects
multi-hit updates to become inefficient. Figure \ref{fig:autocorr} shows the
resulting integrated autocorrelation times \cite{sokal:92a,weigel:10}
$\tau_\mathrm{int}$ of the internal energy of a $128\times 128$ Ising system as a
function of inverse temperature $\beta$ and comparing the CPU calculation with GPU
simulations with multi-hit updates ranging from $k=1$ up to $k = 100$, using tiles of
$T\times T = 16\times 16$ spins. Note that, in the same way as sequential spin
updates in a scalar simulation have different autocorrelation times than purely
random updates, the checkerboard (i.e., GPU) way of updating the spins yields
slightly different autocorrelation times than the sequential update on CPU. This
effect, however, appears to be weak. Apart from the general expected rounded
divergence of autocorrelation times in the vicinity of the critical coupling $\beta_c
= \frac{1}{2}\ln (1+\sqrt{2}) \approx 0.44069$, Fig.~\ref{fig:autocorr} clearly
reveals the relative increase of $\tau_\mathrm{int}$ as $k$ is increased. As a
result, in the vicinity of $\beta_c$ there are two competing effects: increasing $k$
reduces the overhead of loading tiles from global memory and thus decreases spin-flip
times, whereas at the same time it increases autocorrelation times, thus reducing
sampling efficiency. As a result, the autocorrelation time in physical units,
$$
\tilde{\tau}_\mathrm{int} = \tau_\mathrm{int} t_\mathrm{flip},
$$
where $t_\mathrm{flip}$ denotes the real time for a single spin flip, such that $L^2
\tilde{\tau}_\mathrm{int}$ corresponds to the real time for generating an effectively
uncorrelated spin configuration, close to $\beta_c$ has a minimum at intermediate
values of $k$. This is illustrated in Fig.~\ref{fig:autocorr2}, again for a
$128\times 128$ system with tile size $T = 16$, where it is seen that, in the
vicinity of $\beta_c$, multi-hit updates with $k \approx 10$ appear to be optimal.

\subsection{Performance in 2D}

Using the double checkerboard decomposition loading tiles into shared memory and the
RNGs discussed above, an optimized code was first constructed for the case of the
square-lattice Ising model. The sample code, which has undergone significant
further optimizations as compared to the version discussed in Ref.~\cite{weigel:10c},
can be downloaded from the author's web site \cite{weigel:gpu}.

\begin{figure}[tb]
  \centering
  \includegraphics[keepaspectratio=true,scale=0.75,trim=45 48 75 78]{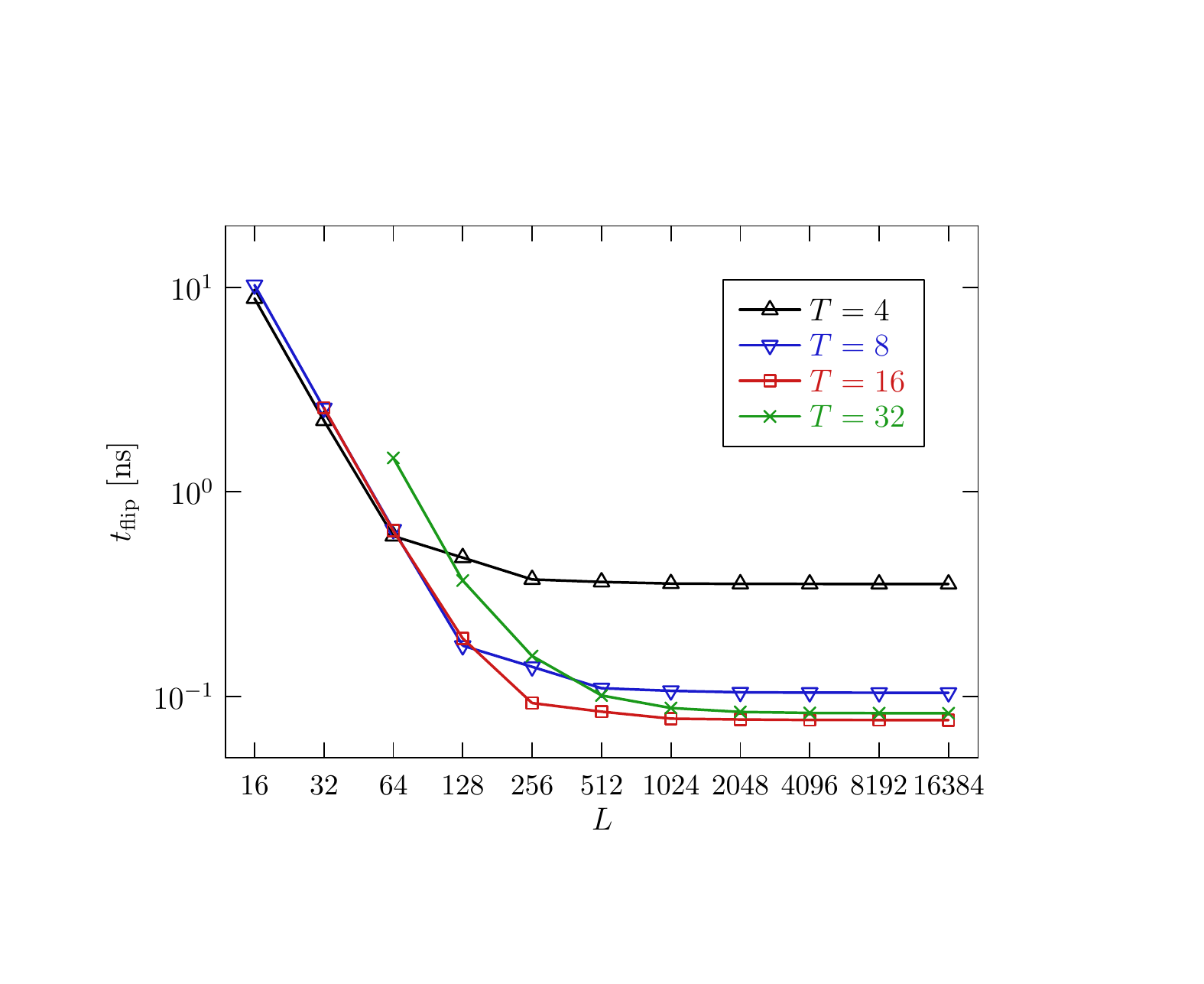}
  \caption{Spin-flip times in nanoseconds for $k = 100$ multi-hit updates of the
    square-lattice Ising model as a function of linear system size $L$. The lines
    correspond to different block sizes $T$.
  }
  \label{fig:blocks}
\end{figure}

It is a well-known trick valid for discrete models to tabulate the exponential
occurring in the acceptance criterion (\ref{eq:metropolis}) in advance and use a
look-up instead of calculating the exponential directly. It appears natural to store
this array in constant memory which is fast through caching \cite{kirk:10}. This
memory, however, is optimized for the different threads of a block synchronously
accessing the {\em same\/} memory element, which is not the typical case here since
different spins usually have different local interaction energies. It is found more
efficient, therefore, to store this table in texture memory which is fast with
different threads accessing different memory locations. While for the CPU code it is
beneficial to save arithmetic cycles by drawing random numbers and looking up the
value of the Boltzmann factor only when $\Delta E > 0$, this is not the case on GPU,
where such an approach results in thread divergence requiring the corresponding
branch to be visited twice. For the code discussed here, I used a single variable of
the high-level language to represent a spin. Performance differences depending on
whether {\tt int} (32 bit), {\tt short int} (16 bit) or {\tt char} (8 bit) variables
are used are only very moderate. The results discussed below are for {\tt int}
variables. A multi-spin coded version is discussed below in Sec.~\ref{sec:spinglass}.

\begin{figure}[tb]
  \centering
  \includegraphics[keepaspectratio=true,scale=0.75,trim=45 48 75 78]{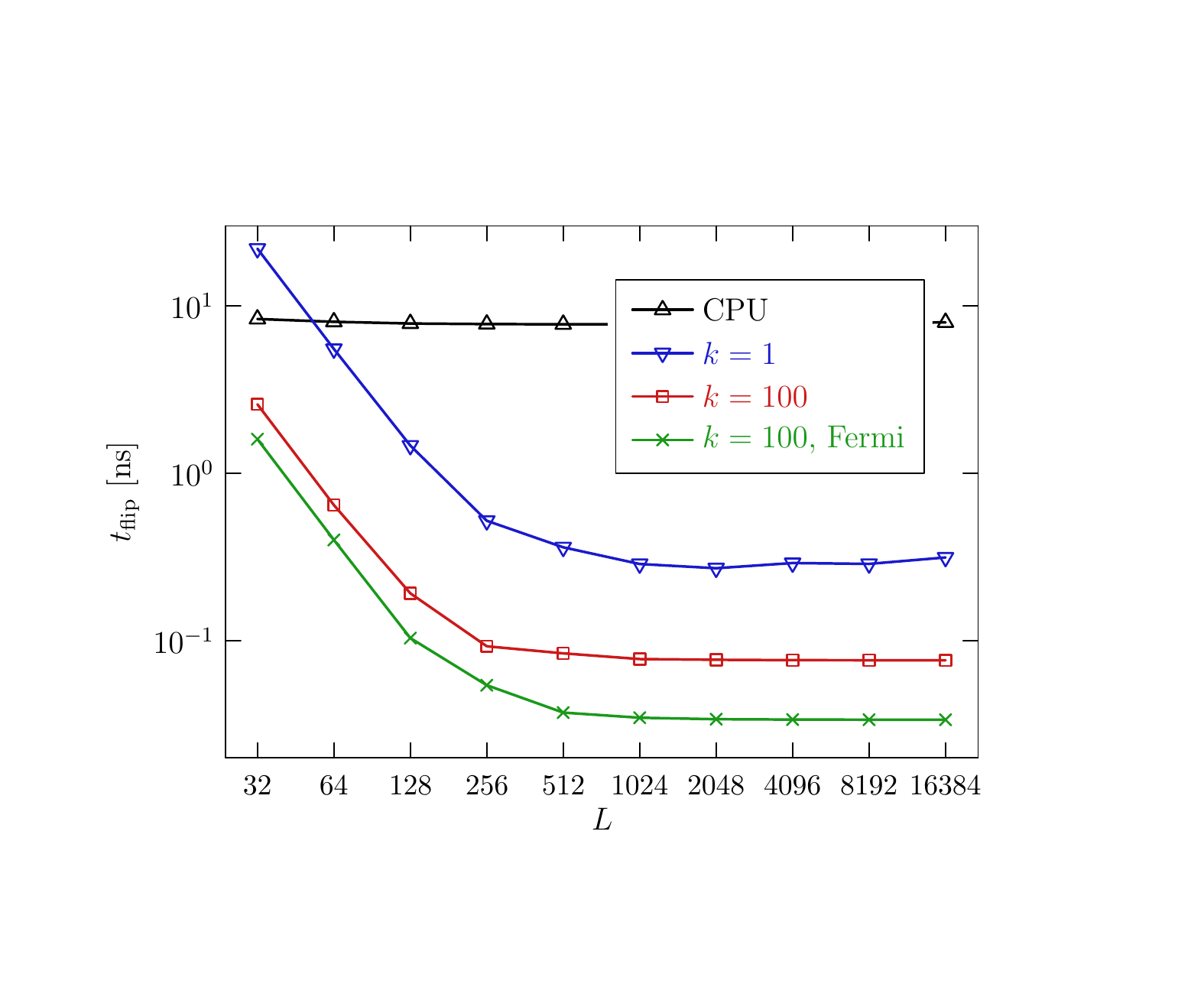}
  \caption{Time per spin-flip in the 2D Ising model on CPU and on GPU with different
    choices of $k$. GPU data are for the Tesla C1060 device apart from the lowest
    curve which is for a GTX 480 card. All data are for tile size $T= 16$ and an
    inverse temperature $\beta = 0.4$ close to, but off the critical coupling. There
    is a weak temperature dependence of spin-flip times due to differences in the
    number of write operations.}
  \label{fig:speedup}
\end{figure}

The optimum choice of the size $T$ of tiles depends on system size and is determined
by the relative resource utilization on the considered GPU device. The corresponding
measured spin-flip times for $k = 100$ shown in Fig.~\ref{fig:blocks} are easily
understood from the parameters of the Tesla C1060 device at hand: tiles should be
large enough to provide a sufficient number of threads in each block. The optimum
load is seen when each multiprocessor is assigned with the maximum number of eight
thread blocks. Since each multiprocessor can accommodate a maximum allowed number of
$1024$ threads, this results in $128$ threads per block. This number is reached for
tiles of size $16\times 16$ spins (remember that each thread updates two sites). This
choice turns out to be optimal for lattice sizes $L\ge 256$. For smaller lattices, on
the other hand, $T = 16$ does not provide enough blocks to load all 30
multiprocessors of the device which requires $B^2/2 = (L/T)^2/2 \ge 30$ or $T\le 8$
for $L = 128$ and $T\le 4$ for $L = 64$. Since $T = 4$ is the smallest tile size
considered, this is also optimal for the smallest lattice sizes $L = 16$ and $L =
32$. For smaller values of $k$, the performance of global memory accesses comes into
play, which favors larger tile sizes due to the resulting improvements in coalescence,
such that $T = 16$ turns out to be optimal for any $L\ge 32$ there. Test simulations
for the Fermi architecture on a GTX 480 card lead pretty much to the same conclusions:
there, the maximum number of resident threads per multiprocessor has been increased
to 1536, but this increase is too small to lift the optimum tile size to $T = 32$
such that, again, $T=16$ is optimal apart from simulations on the smallest lattices.

\begin{figure}[tb]
  \centering
  \includegraphics[keepaspectratio=true,scale=0.75,trim=45 48 75 78]{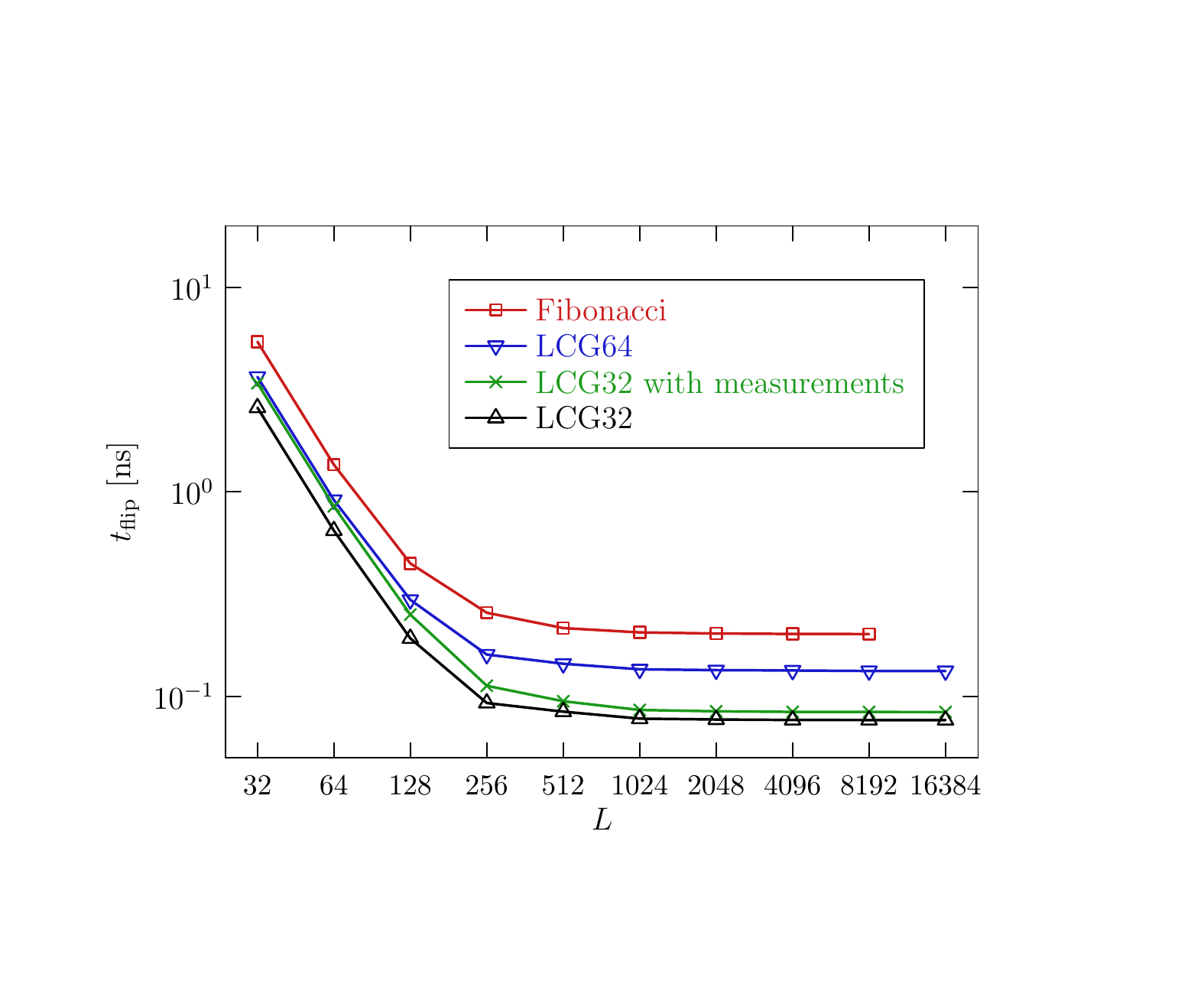}
  \caption{Spin-flip times in nanoseconds for the 2D Ising GPU code employing
    different random number generators and showing the effect of including energy
    measurements. All runs for $k = 100$.
  }
  \label{fig:speedup2}
\end{figure}

Figure \ref{fig:speedup} shows the performance data for the final code as compared to
the scalar CPU implementation. As a reference CPU I used an Intel Core 2 Quad Q9650
at 3.0 GHz, which is one of the fastest available CPUs for single-threaded
applications. The performance of the latter is essentially independent of system size
with about $8$~ns per spin flip. The maximum performance on GPU is reached only for
system sizes $L \gtrsim 4096$ with about $0.27$~ns per spin flip for $k = 1$ and
$0.076$~ns per spin flip for $k = 100$, resulting in speed-up factors ranging from
$30$ to $105$. The Fermi architecture results in an extra speed-up of about $2.3$
compared to the Tesla C1060 device, leading to a maximum performance of $0.034$~ns
per spin flip at $k = 100$ or an overall speedup factor of $235$ (the extra speedup
for $k=1$ is even slightly larger, probably due to the reduced overhead of kernel
calls for the Fermi architecture). Comparing to the previous implementation in
Ref.~\cite{preis:09}, one notes that, probably due to inappropriate cache alignment,
the (single-spin coded) CPU code used there is rather inefficient, such that, e.g.,
for a $L=4096$ system the code used here is about $10$ times faster than that of
Ref.~\cite{preis:09}
and $5$ times faster than the results quoted in Ref.~\cite{block:10},
leading to a rather unfavorable comparison for the CPU based systems in these
works. Their multi-spin coded CPU version is less than a factor of two more efficient
than the single-spin coded implementation used here (see below in
Sec.~\ref{sec:spinglass} for a multi-spin coded version of the program used
here). The maximum performance of the GPU implementation of Ref.~\cite{preis:09} with
$0.7$ ns on the Tesla C1060 is about $2.6$ times less than the present $k = 1$ code
and a factor of $9$ slower at $k = 100$. The significant speedup at $k = 1$ is due to
the combined effect of the avoidance of bank conflicts in computing the local
interaction energy, the fact that periodic boundaries do not need to be taken account
with expensive modulo operations once the tile including its boundary layer has been
loaded into shared memory, and the efficient implementation of the Metropolis
criterion using texture fetches. The multi-spin coded GPU code (shared memory
version) of Ref.~\cite{block:10} is about of the same performance as our single-spin
coded version at $k = 1$.

\begin{figure}[tb]
  \centering
  \includegraphics[keepaspectratio=true,scale=0.75,trim=45 48 75 78]{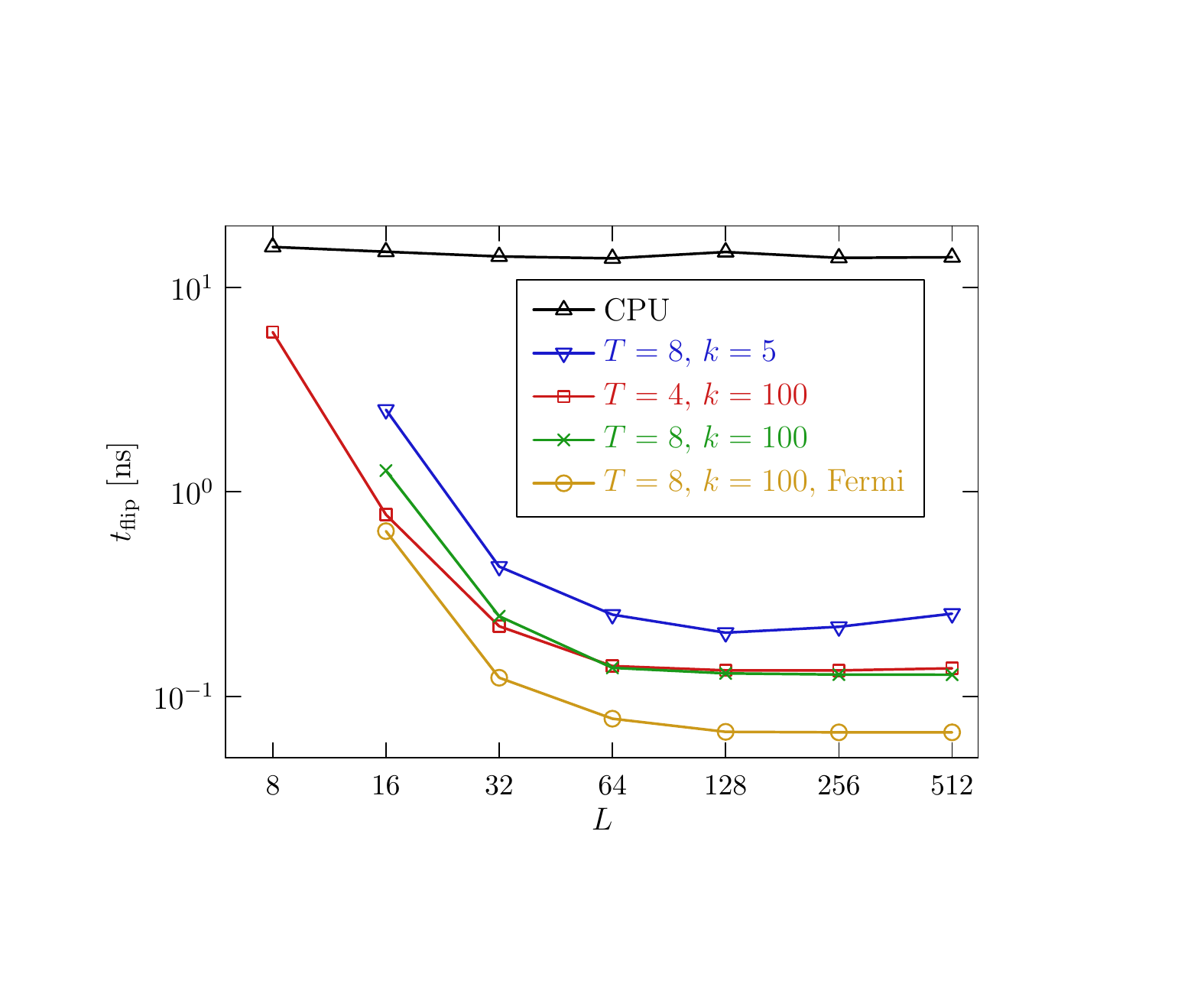}
  \caption{Timings per spin flip for the 3D Ising CPU code on the Tesla C1060 and
    GTX 480 (Fermi) devices as compared to the reference CPU implementation.
  }
  \label{fig:speedup3D}
\end{figure}

In a practical application measurement cycles are required on top of the updates, of
course. Energy measurements were here realized with tracking the local energy changes
at each spin flip and combining them in a final parallel reduction step after the
update. At $k = 100$, corresponding to only one measurement per one hundred lattice
sweeps, the overhead is small, of the order of 10\%, see the data collected in
Fig.~\ref{fig:speedup2}. At $k = 1$, the overhead is about 20\%. In view of the
potential problems with the quality of random numbers discussed above in
Sec.~\ref{sec:rng}, it is worthwhile to check the influence of the different speeds
of RNGs discussed above on the overall performance of the Ising simulation code. In a
simulation as simple as the Metropolis update of the Ising model, a substantial
fraction of time is spent for generating random numbers. Therefore, it is no surprise
that it is found that the simulation is slower by a factor of $1.73$ when using the
LCG64 instead of the LCG32 generator and by a factor of $2.63$ when using the lagged
Fibonacci generator ($k = 100$, $L = 16\,384$). In view of the good results for the
randomly initialized LCG32 reported in Sec.~\ref{sec:rng}, it is probably acceptable
to use it for the considered model in view of its good performance. For other
applications, however, this issue would need to be revisited.

\subsection{Performance in 3D}

The generalization of the GPU code from two to three dimensions is straightforward,
and only the part responsible for the correct bookkeeping in loading a tile into
shared memory becomes more tedious to formulate. The range of allowable tile sizes is
somewhat more restricted by the limitation of the available GPUs to 512 threads per
block (resp.\ 1024 for Fermi). With the current setup, therefore, the only tile sizes
that can be reasonably considered are $T=4$ and $T=8$, if one restricts oneself to
powers of two in system as well as in tile sizes. More general and, perhaps,
versatile, setups could be reached by relaxing the power-of-two condition or
considering tiles with non-unit aspect ratios, which might lead to additional slight
increases in performance. For the sake of simplicity, however, I here refrain from
introducing such complications.


The timing results for the 3D code are summarized in Fig.~\ref{fig:speedup3D}. The
performance of the CPU code is essentially independent of lattice size within the
considered range of $L=8$ up to $L = 512$, arriving at around $14$ ns per spin flip
on the reference CPU used here, which is slower than the 2D code by a bit more than
the factor of $6/4 = 3/2$ expected merely from the increased number of nearest
neighbors to be included in energy calculations. For the GPU code, we find rather
comparable performance from $T=4$ and $T=8$ with a slight advantage for the latter
for lattice sizes exceeding $L=64$. The maximum performance on the Tesla C1060 is
around $0.13$ ns (at $k=100$), about by the expected factor of $2/3$ less than in two
dimensions. The Fermi card is faster by another factor of $1.9$, resulting in a peak
performance of $0.067$ ns per spin flip. As a consequence, the observed maximal
speed-up factors are $110$ for the C1060 and $210$ for the GTX 480. These speed-ups
are very close to those observed in two dimensions, such that the relative efficiency
of the CPU and GPU implementations are practically identical in two and three
dimensions.

\section{Continuous spins\label{sec:heisen}}

A natural generalization of the Ising model is to systems with continuous degrees of
freedom, i.e., O($n$) spin models with Hamiltonian
\begin{equation}
  \label{eq:On_hamiltonian}
  {\cal H} = -J\sum_{\l i,j\r}\bm{s}_i \cdot \bm{s}_j,\;\;\;|\bm{s}_i| = 1,
\end{equation}
which are, most often, more realistic for the description of real magnetic
materials. For implementations of simulation algorithms on GPU, this offers the
opportunity to test the relative performance of floating-point operations, an
application for which GPUs are highly optimized by design, since their original
purpose in rendering graphics in technical terms to a large percentage reduces to
manipulations of matrices of floating-point numbers.

Since for the original graphics purposes calculations in double precision usually are
not relevant, traditionally GPU devices have only offered single-precision
arithmetic. Although double precision calculations could be emulated in software,
native support for double precision arithmetic was only introduced with the GT200
architecture, but it came with a massive performance penalty, double precision being
around eight times slower than single precision. Only with the recent Fermi
architecture, this drawback has been partially removed, double precision calculations
there being only around a factor of two slower than single precision
arithmetic. Sufficient precision is certainly important for numerical algorithms that
have a potential of accumulating rounding and finite-precision errors such as, e.g.,
molecular dynamics simulation codes. For Monte Carlo simulations, on the other hand,
it is rather obvious that due to the random sampling rounding errors and numerical
accuracy are much less of an issue, and it might be acceptable to represent the
degrees of freedom in single precision and use double precision only for accumulated
quantities in measurements. Implementations in single and in double precision are
benchmarked here, also from the point-of-view of the resulting accuracy.

\begin{figure}[tb]
  \centering
  \includegraphics[keepaspectratio=true,scale=0.75,trim=45 48 75 78]{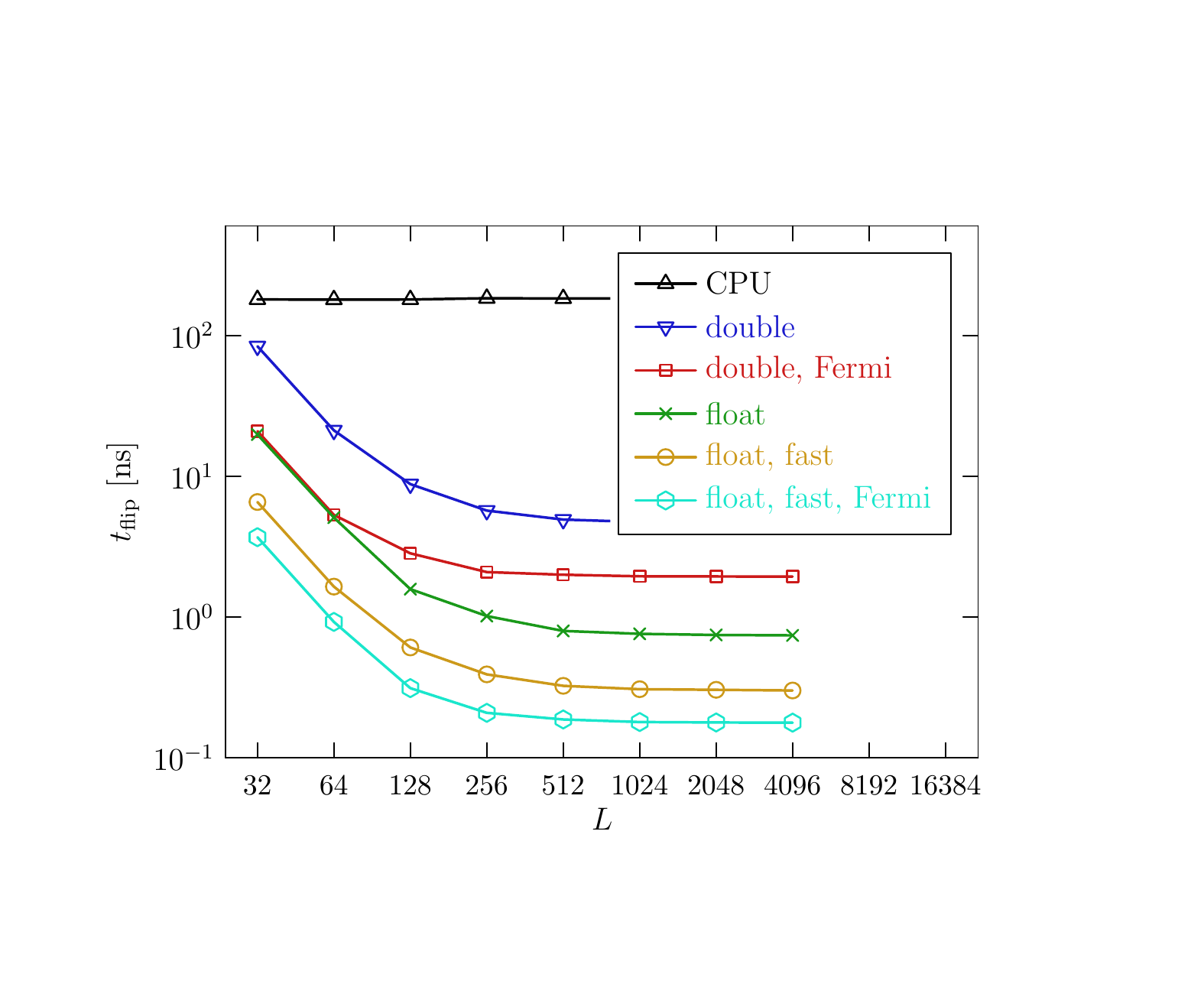}
  \caption{Timings per spin flip for the 2D Heisenberg model simulations on the Tesla
    C1060 and GTX 480 (Fermi) devices as compared to the reference CPU
    implementation. The data are for $L=4096$ and $k=100$.
  }
  \label{fig:heisen}
\end{figure}

In addition to the question of the number of digits, the floating-point
implementation on CUDA devices is not completely compliant with the IEEE-754
floating-point standard which is otherwise practically universally accepted for
current computing platforms. In particular, devices prior to the Fermi series do no
support the so-called denormalization feature (for single-precision calculations)
that ensures sufficient accuracy for numbers close to zero. Furthermore, due to
device-specific implementations, the accuracy of operations such as division and
square root is sometimes less than the $0.5$ units in the last place (ULP) prescribed
by the IEEE standard. Additionally, there are some less dramatic deviations such as
the absence of a signaling of floating-point exceptions, the lack of configurable
rounding modes and others.  The mentioned limitations are removed with the advent of
Fermi cards, which can be configured to almost completely adhere to the IEEE
standard. Regarding transcendental functions such as exponential, sine and logarithm,
device-specific and highly performant, but less accurate, versions are provided in
the CUDA framework in addition to the standard C library implementations that are
also available \cite{cuda}. (Note, however, that these highly performant
implementations are only available in single-precision variants.) Apart from these
questions of adherence to the floating-point standard and whether single-precision
calculations are sufficient for the purpose at hand, one must keep in mind that
parallel algorithms in themselves might lead to finite-precision effects which are
different from those seen in serial algorithms. If for instance, a binary-tree
algorithm is used to perform a parallel reduction over an array of values to evaluate
their sum (as in the energy measurement code of the algorithms discussed here), the
result will be slightly different to that found in a serial calculation since in
finite precision addition is not commutative. Such effects can be reduced, however,
with appropriate adaptations, for instance sorting the array by size prior to the
reduction, such that values of similar magnitude are added first. Such considerations
are not specific to GPUs, of course, but need to be taken into account whenever
parallel algorithms come into play.

To study the effects of these particularities of the massively parallel GPU
architecture on the performance and accuracy of a Monte Carlo simulation, a
Metropolis single-spin update simulation code was benchmarked for the case of the
Heisenberg model. For the sake of simplicity, a new spin orientation is proposed
uniformly at each update step, independent of the previous orientation of the
spin. In practice, this leads to poor acceptance rates at low temperatures which
could be improved by proposing small modifications instead of uniform
re-orientations. For the benchmarks considered here, however, this question is
irrelevant, since local reorientations lead to almost identical results for the
relative performance of the different implementations considered. In the standard
implementation, drawing a random unit vector in three dimensions requires two random
numbers and a total of five trigonometric function calls. In addition, there are $6d$
multiply-and-add operations for calculating the scalar products in determining the
local energy change in an attempted flip for simulations on a $d$-dimensional
(hyper-)cubic lattice. Compared to the Ising system, the exponentials in the
Metropolis criterion (\ref{eq:metropolis}) cannot be tabulated in advance, leading to
an extra special-function call per attempted update. As a result of these
differences, an update step for the Heisenberg model is significantly heavier with
arithmetic operations than an Ising update which, however, from the point-of-view of
GPU computing is a rather favorable situation since this reduces the probability of
memory bandwidth limitations. (Note, however, that Heisenberg updates also require
larger memory transfers since each spin variable requires three floating-point
numbers.)

\begin{table}
  \centering
  \begin{tabular}{llt{2}t{0}} \hline
    \multicolumn{1}{c}{device} & \multicolumn{1}{c}{mode} & \multicolumn{1}{c}{$t_\mathrm{flip}$ [ns]} & \multicolumn{1}{c}{speedup} \\ \hline
    CPU (Intel Q9650) & float or double & 183.16                                 & 1     \\ \hline
    & float         &   0.74                                 & 247   \\
    Tesla C1060 & float, fast\_math   &   0.30                                 & 608   \\
    & double        &   4.66                                 & 39    \\ \hline
    & float         &   0.50                                 & 366   \\
    GTX 480 & float, fast\_math   &   0.18                                 & 1029   \\
    &double        &   1.94                                 & 94    \\ \hline
  \end{tabular}
  \caption{Times per spin flip in ns for various implementations of single-spin flip
    simulations of the 2D Heisenberg model and speedup factors as compared to the CPU
    reference implementation. All data are for multi-hit updates with $k=100$ and
    system size $L=4096$.}
  \label{tab:heisen}
\end{table}

Figure \ref{fig:heisen} compares the resulting spin-flip times for various GPU and
CPU implementations and a system size of $4096\times 4096$ spins. The CPU
implementation has exactly the same performance for single and double precision
implementations, which is plausible since most arithmetic units in current CPUs are
double precision anyway. For single-precision calculations, the use of the intrinsic,
fast implementations of the special functions results in a factor of 2.5 speedup of
the entire code, while the use of double variables for representing spins results in
an about sixfold slow-down for the Tesla C1060 and a fourfold slow-down for the Fermi
card. Overall speed-up factors are rather impressive, ranging from 39 for the
double-precision version on the Tesla C1060 and 1029 for the single-precision code
with fast special functions on the Fermi board, cf.\ the data collected in
Tab.~\ref{tab:heisen}. It should be noted, of course, that reduced-precision
implementations of special functions are, in principle, also possible on CPU
architectures, such that the extra speed-up gained by the use of these functions is
not specific to using GPUs.

\begin{figure}[tb]
  \centering
  \includegraphics[keepaspectratio=true,scale=0.75,trim=45 48 75 78]{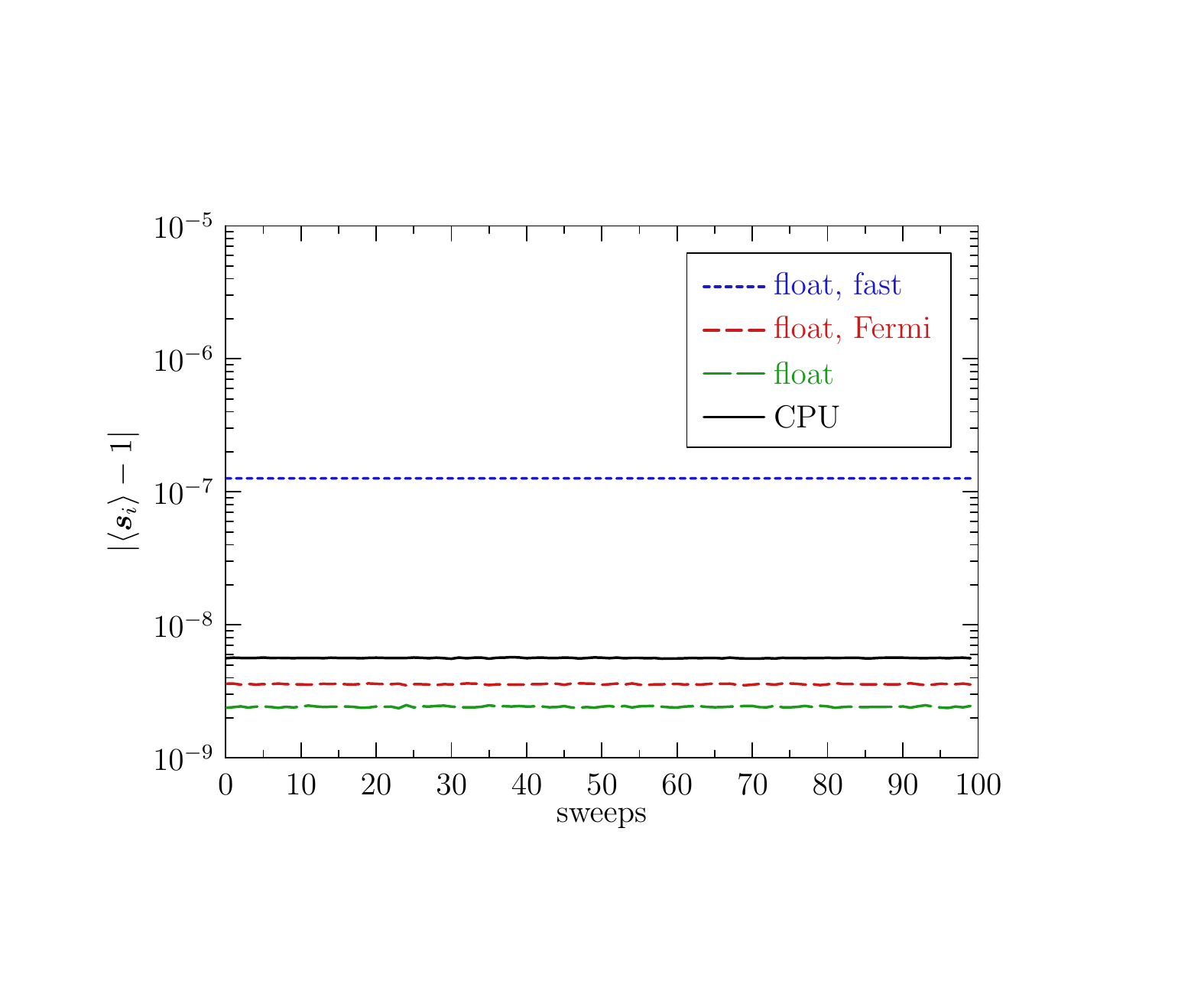}
  \caption{Average deviation of the spin normalization from unity in Heisenberg model
    simulations employing single-precision floating point arithmetic on CPU and on
    GPU as a function of the number of lattice sweeps.
  }
  \label{fig:stability_heisen}
\end{figure}

To see whether numerical stability and finite precision are relevant for the
calculations considered here, deviations from the required normalization $|\bm{s}_i|
= 1$ have been monitored for the different implementations. As can be seen from the
results presented in Fig.~\ref{fig:stability_heisen}, a drift of normalization is not
an issue here. As expected, deviations of $|\bm{s}_i|$ from unity are of the order of
$10^{-8}$ for single-precision implementations uniformly for CPU as well as GPU
implementations. Maybe surprisingly, deviations are even somewhat smaller for the GPU
codes than for the CPU implementation. Using the fast intrinsic implementations of
special functions, on the other hand, reduces the accuracy by about one digit, such
that deviation are of the order of $10^{-7}$ there. Estimates of the magnetization,
internal energy and specific heat extracted from the simulation runs also have been
compared between the different implementations and, at the given level of statistical
accuracy, no significant deviations have been observed. (Note that reductions in the
measurement cycles have been implemented in double precision.) It appears justified,
therefore, to claim that for spin model simulations a representation of the spin
variables in single precision, while still doing measurements in double precision,
does not introduce systematic deviations of relevance as compared to statistical
fluctuations. The same seems to be valid even when using the fast intrinsic, reduced
accuracy implementations of the special functions. In view of the dramatic speed-ups
observed in particular for the ``float, fast\_math'' implementation, cf.\
Tab.~\ref{tab:heisen}, this appears encouraging.

\section{Spin glass\label{sec:spinglass}}

Computationally particularly demanding problems in the field of spin systems occur
for models with quenched disorder \cite{young:book}. The nature of fluctuations
between disorder realizations typically requires averages over at least several
thousand disorder configurations to enable reliable results. For the case of spin
glasses, where disorder is accompanied by the presence of competing interactions,
each disorder configuration exhibits extremely slow dynamics close to or below the
spin-glass transition. Since no effective cluster algorithms are available for spin
glasses (with the exception of systems in two dimensions \cite{houdayer:01}),
simulations are restricted to single-spin updates, and the combined effect of
disorder average and slow relaxation results in rather excessive computational
requirements. It is not surprising, therefore, that a number of special-purpose
computers have been constructed for such problems, the most recent being the Janus
machine based on FPGAs \cite{belleti:09}.

\begin{figure}[tb]
  \centering
  \includegraphics[keepaspectratio=true,scale=0.75,trim=45 48 75 78]{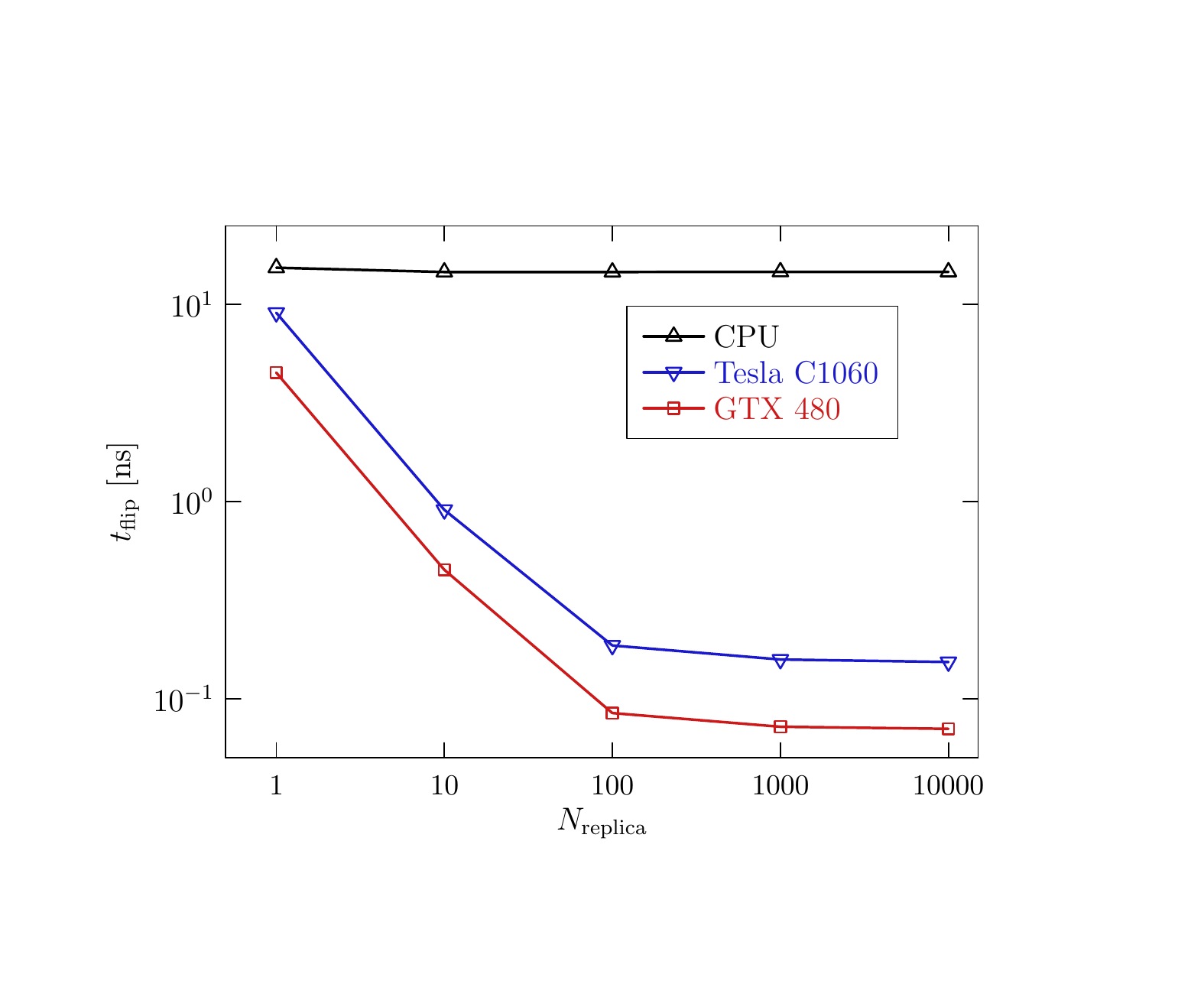}
  \caption{Spin-flip times in ns for simulating $N_\mathrm{replica}$ independent
    disorder realizations of a $16\times 16$ Ising spin glass system on the GT200
    and Fermi architectures as compared to the CPU implementation.
  }
  \label{fig:spinglass}
\end{figure}

The trivial parallelism inherent in the disorder average and the necessary
restriction to local spin updates appears to render spin-glass problems ideal
candidates to profit from the massively parallel architecture of GPUs. To see to
which degree such hopes are borne out in real implementations, I studied the
short-range Edwards-Anderson Ising spin glass model with Hamiltonian
\begin{equation}
  \label{eq:spinglass_hamiltonian}
  {\cal H} = -\sum_{\l i,j\r} J_{ij} s_i s_j,\;\;\;s_i = \pm 1,
\end{equation}
where the exchange couplings $J_{ij}$ are quenched random variables, here drawn from
a symmetric bimodal distribution $P(J_{ij}) =
\frac{1}{2}\delta(J_{ij}-J)+\frac{1}{2}\delta(J_{ij}+J)$. Compared to the
ferromagnetic Ising model \eqref{eq:ising_hamiltonian}, the disordered version
\eqref{eq:spinglass_hamiltonian} requires the values of the couplings $J_{ij}$ to be
taken into account when calculating the energy change of a spin flip such that an
additional data structure for the couplings becomes necessary. For the tiled
algorithm described in Sec.~\ref{sec:checker} above, the values of $J_{ij}$ inside of
each tile need to be loaded into shared memory in addition to the spin
field. Furthermore, a number $N_\mathrm{replica}$ of completely independent instances
of the spin system with different coupling values can be simulated in parallel. These
different instances are mapped to different thread blocks in the GPU
architecture. Apart from these changes, the algorithm is identical to the
implementation described for the ferromagnetic model.

\begin{figure}[tb]
  \centering
  \includegraphics[keepaspectratio=true,scale=0.75,trim=45 48 75 78]{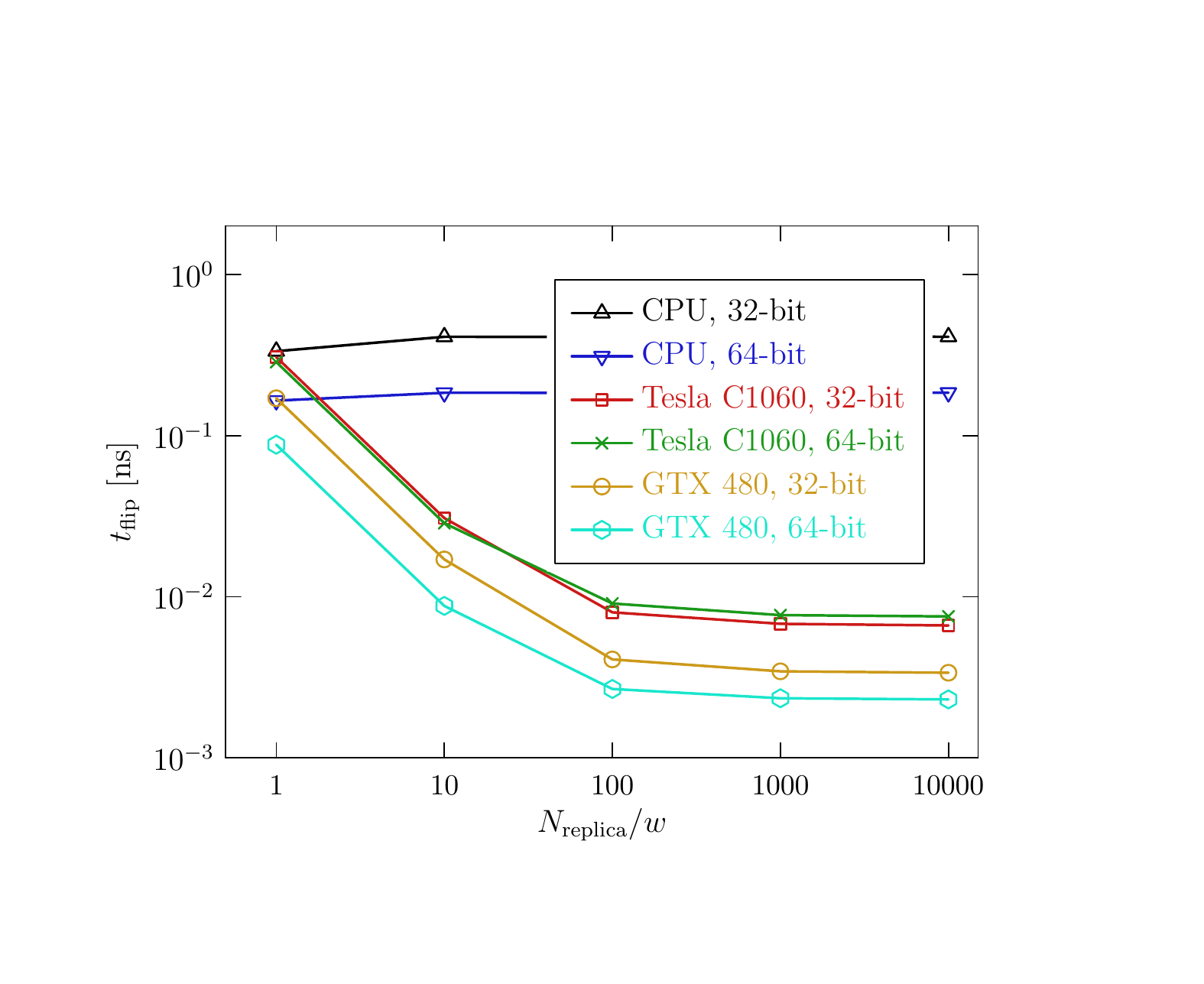}
  \caption{Spin-flip times in ns for simulations of the Ising spin-glass using
    asynchronous multi-spin coding with either $w=32$ or $w=64$ bit variables as a
    function of the number of disorder realizations. Simulations are for a $16\times
    16$ system, comparing CPU and GPU codes on various devices.}
  \label{fig:spinglassmsc}
\end{figure}

The spin-flip times resulting from this implementation are displayed in
Fig.~\ref{fig:spinglass} as a function of the number $N_\mathrm{replica}$ of disorder
realizations simulated simultaneously. The reference CPU implementation performs at
around $14.6$ ns per spin flip on the Intel Q9650 CPU used here, essentially
independent of the number of realizations as those are merely worked on in sequence
there. The GPU code reaches a maximum performance of $0.15$ ns on the Tesla C1060 and
$0.07$ ns on the Fermi card, respectively, corresponding to speed-ups of $95$ (C1060)
and $208$ (GTX 480). These speed-ups are practically identical to the maximum
speed-ups observed for the ferromagnetic model, whereas in absolute terms due to the
extra calculations related to the non-uniform couplings the CPU and GPU
implementations for the spin-glass model are about a factor of two slower than the
ferromagnetic codes.

It is well known that storing the inherently one-bit information of the orientation
of an Ising spin in a (typically) 4-byte integer variable is a wasteful operation and
packing several spin variables in each machine word can result in significant
performance improvements \cite{zorn:81}. While for the case of ferromagnetic models
the only possible implementation needs to pack a number of spins from the same
lattice into one word of $w$ bits, leading to a somewhat involved sequence of bit-wise
operations becoming necessary in order to perform the Metropolis updates of the coded
spin in parallel \cite{zorn:81} (``synchronous'' multi-spin coding
\cite{belleti:09}), the situation in the presence of quenched disorder implies that
one can also code $w$ spins in the same spatial location of {\em independent\/} Ising
lattices corresponding to different disorder configurations in one word
(``asynchronous'' multi-spin coding \cite{belleti:09}). This setup allows for very
efficient implementations using only a few bit-wise operations. Depending on the type
of integers used, currently $w=32$ or $w=64$ spins can be represented in one
variable. In parallel, 32 or 64 values of bimodal couplings of different realizations
are stored in another field of integers. The resulting performance characteristics
are shown in Fig.~\ref{fig:spinglassmsc}. The CPU code performs at $0.41$ ns per spin
flip for $w=32$ and a significantly better $0.18$ ns for $w=64$, corresponding to a
speedup of $35$ and $79$ as compared to the single-spin coded CPU version for $w=32$
and $w=64$, respectively\footnote{Note that here I am using the {\em same\/} random
  number to update each of the $w=32$ or $w=64$ spins, which appears justified in
  view of the statistical independence of the coupling realizations. This setup was
  previously tested for the Ising spin glass in three dimensions and no significant
  correlation effects were observed \cite{hasenbusch:08}.}. The GPU implementation
performs at around $6.6$ ps for 32-bit and $7.5$ ps for 64-bit on the Tesla C1060
card. The rather disappointing result for the $w=64$ version shows that the GT200
architecture is not optimized for $64$-bit operations. On the GTX 480 Fermi card, on
the other hand, the $64$-bit multi-spin coded implementation runs as fast as $2.3$ ps
per spin flip as compared to $3.4$ ps for the $32$-bit version. The resulting optimal
speed-ups compared to the CPU code are hence around $30$ for the Tesla C1060 and
around $80$ for the GTX 480. For the Fermi card, the resulting code runs around $55$
times faster than the (synchronous) multi-spin coded simulation of the ferromagnetic
model discussed in Ref.~\cite{block:10}. It is worthwhile to note that Janus performs
at around $16$ ps per spin flip on the 3D Ising spin glass
\cite{belleti:09}. Circumstances forbid a clear-cut comparison here, however, since
the Janus code uses synchronous multi-spin coding and a somewhat better random-number
generator. Still, it appears fair to say that an appropriate GPU implementation seems
to play in the same league as the special-purpose computer Janus at a much smaller
investment in capital and human resources.

\section{Parallel tempering\label{sec:tempering}}

A rather powerful technique for the simulation of systems with complex free energy
landscapes, including spin glasses and bio-polymers, is given by the parallel
tempering or replica exchange algorithm \cite{geyer:91,hukushima:96a}. There, a
number of identical copies of a system are simulated at a set of different, closely
neighboring temperatures. In addition to any chosen intra-replica update, such as the
single spin flips according to the Metropolis criterion \eqref{eq:metropolis}, at
fixed intervals an interchange of configurations (typically) simulated at neighboring
temperatures is attempted and accepted with the generalized Metropolis probability
\begin{equation}
  \label{eq:parallel_tempering}
  p_\mathrm{acc}(\{s_i\},\beta \mapsto \{s_i'\},\beta') = \min\left[1,\,e^{\Delta \beta\Delta E}\right],
\end{equation}
where $\Delta E = E(\{s_i'\})-E(\{s_i\})$ and $\Delta \beta = \beta'-\beta$. With a
properly chosen set of temperatures \cite{katzgraber:06,bittner:08,hasenbusch:10},
this additional update allows for configurations with slow dynamics at low
temperatures to successively diffuse to high temperatures, decorrelate there and
finally return back to the low-temperature phase, likely arriving in a different
valley of a complex free-energy landscape than where it started from.

\begin{figure}[tb]
  \centering
  \includegraphics[keepaspectratio=true,scale=0.75,trim=45 48 75 78]{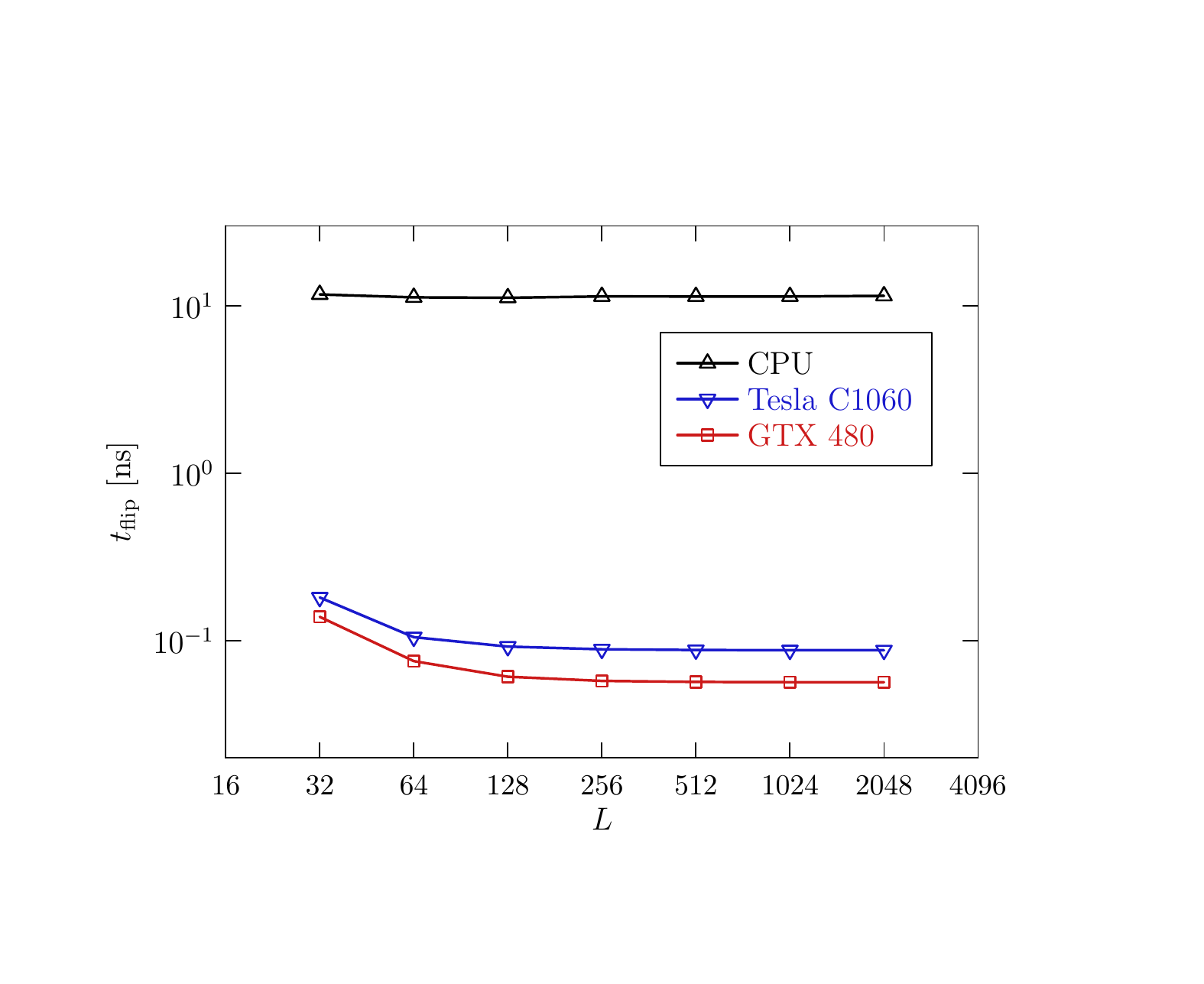}
  \caption{Spin-flip times for a parallel-tempering simulation of the 2D Ising model
    as a function of system size. Simulations were performed on the range of inverse
    temperatures $\beta = [0.1,0.15]$ with $n = 20$ replicas with one exchange
    attempt per 100 sweeps of spin flips.
  }
  \label{fig:parallel}
\end{figure}

Through its inherent parallelism of intra-replica updates, this scheme appears to be
well suited for the massively parallel architecture of current GPUs. It is tested
here in a reference implementation for the ferromagnetic Ising model along the lines
of the previously discussed single spin-flip code. The additional copies of the
system are mapped to additional thread blocks running in parallel. Since the
configurational energy is calculated online from the energy changes of single spin
flips via a binary-tree reduction algorithm on tiles, it is automatically available
after lattice sweeps and hence its determination does not incur any extra
overhead. In the current implementation, replica exchange steps are performed on CPU
since the computational effort is low and synchronization between thread blocks is
required. As usual, instead of exchanging configurations between replicas only the
corresponding temperatures are swapped. The Boltzmann factors according to
(\ref{eq:metropolis}) can still be tabulated, and are now implemented on GPU as
fetches from a two-dimensional texture. A successful replica exchange then requires
an update of the texture which is easily possible in the current setup as the
exchange moves are carried out on CPU.

The benchmark results of the parallel tempering simulation of the 2D Ising
model on GPU and CPU is shown in Fig.~\ref{fig:parallel}. I chose to use $n = 20$
replicas at equally spaced inverse temperatures in the interval $\beta =
[0.1,0.15]$. In a real-world application one would probably want to use a more
elaborate scheme for choosing these temperatures \cite{katzgraber:06,bittner:08}, but
these questions do not have any bearing on the benchmark calculations performed
here. As is clearly visible, the presence of additional copies of the system leads to
a much better resource utilization for smaller system sizes than that observed for
the single-spin flip simulations. Hence, significant speed-ups are observed already
for small systems. The CPU code performs at a constant $11$ ns per combined spin-flip
and replica exchange move (mixed at a ratio of one exchange move per one hundred
lattice sweeps). The GPU code arrives at a maximum of around $0.089$ ns for the Tesla
C1060 and $0.057$ ns for the GTX 480 at $L = 2048$. The speed-up reaches up to $130$
(C1060) resp.\ $200$ (GTX 480) at $L = 2048$, but is already $109$ (C1060) resp.\
$151$ (GTX 480) for the $L = 64$ system. As for disordered systems due to the severe
slowing down usually only rather small systems can be studied, good performance of
the code is crucial in this regime. Increasing the number of exchange moves to one in
ten lattice sweeps reduces the maximum performance of the GPU implementation somewhat
to $0.11$ ns (C1060) and $0.070$ ns (GTX 480), respectively.

\section{Conclusions\label{sec:concl}}

Current GPUs have a significant potential for speeding up scientific calculations as
compared to the more traditional CPU architecture. In particular, this applies to the
simulation of systems with local interactions and local updates, where the massive
parallelism inherent to the GPU design can work efficiently thanks to appropriate
domain decompositions. The simulation of lattice spin systems appears to be a
paradigmatic example of this class as the decomposition remains static and thus no
significant communication overhead is incurred.  Observing substantial speed-ups by
factors often exceeding two orders of magnitude as for the case studies reported here
requires a careful tailoring of implementations to the peculiarities of the
considered architecture, however, in particular paying attention to the hierarchic
organization of memories (including more exotic ones such as texture memory), the
avoidance of branch divergence and the choice of thread and block numbers
commensurate with the capabilities of the cards employed. For achieving good
performance, it is crucial to understand how these devices hide the significant
latency of accessing those memories that do not reside on die through the interleaved
scheduling of a number of execution threads significantly exceeding the number of
available computing cores. It is encouraging that employing such devices with the
rather moderate coding effort mediated by high-level language extensions such as
NVIDIA CUDA or OpenCL updating times in the same ballpark as those of special purpose
machines such as Janus \cite{belleti:09} with a vastly higher development effort can
be reached.

A regularly uttered objection against the systematic use of GPUs for scientific
computing criticizes them as being a possibly too special and exotic architecture
with unknown future availability and architectural stability as compared to the
traditional and extremely versatile x86 CPU design. While there is certainly some
truth to this argument, there can be no doubt about the fact that massive parallelism
is not only the present state of the GPU architecture, but also the future of CPU
based computing. As of this writing, Intel CPUs feature up to eight cores and AMD
chips up to twelve cores per die, the corresponding road-maps projecting even
significantly more cores in the foreseeable future. Supercomputers will soon count
their number of cores in the millions. Due to this development, serial computing will
not remain a viable option for serious computational science much longer. Much effort
will need to be invested in the years to come into solving scientific problems
employing a {\em massive\/} number of parallel computational units. In this respect,
GPU computing, apart from currently being more efficient for many problems in terms
of FLOP/s per Watt and per Euro than CPU based solutions, is rather less exotic than
pioneering.

An ideal application for GPU computing in the field of the simulation of spin systems
appear to be disordered systems, where cluster algorithms are in general not
efficient and a natural parallelism occurs from the quenched average over disorder,
possibly combined with the parallel tempering algorithm. Using asynchronous
multi-spin coding for the Ising spin glass, spin flip times down to $2$ ps can be
achieved. Systems with continuous spins are particularly well suited for GPU
deployment, since one finds a relatively stronger overhead of arithmetic operations
over memory fetches as compared to systems with discrete spins there. For the
Heisenberg model, speed-ups up to a factor of $1000$ can be obtained when making
use of the highly optimized special function implementations in single
precision. While these examples of single-spin flip simulations look rather
promising, it is clear that other well-established simulation algorithms for spin
systems are less well suited for the parallelism of GPUs, including the cluster
algorithms in the vicinity of critical points, where spin clusters spanning the whole
system need to be identified, or multi-canonical and Wang-Landau simulations, which
require access to the current values of a global reaction coordinate (such as, e.g.,
the energy or magnetization) for each single spin flip, thus effectively serializing
all spin updates. It is a challenge for future research to find implementations or
modifications of such algorithms suitable for massively parallel computers
\cite{weigel:10b}.

\section*{Acknowledgments}

I am indebted to T.~Yavors'kii for a careful reading of the manuscript. Support by
the DFG through the Emmy Noether Programme under contract No.\ WE4425/1-1 and by the
Schwerpunkt f\"ur rechnergest\"utzte Forschungsmethoden in den Naturwissenschaften
(SRFN Mainz) is gratefully acknowledged.







\bibliographystyle{model1-num-names}


\end{document}